\documentclass[12pt]{iopart}
\usepackage{graphicx}

\newcommand{\bra}[1]{\langle #1 |}			
\newcommand{\ket}[1]{| #1 \rangle}
\newcommand{\proj}[2]{\ket{#1}\bra{#2}}

\newcommand{\EP}{(el)} 
\newcommand{\PP}{(pol)} 

\newcommand{\ham}{H}
\def\figHc{5cm}
\def\figWc{6.7cm}

\def\figW{8cm}

\begin{document}

\title[Charge transfer and coherence of tunnelling system coupled to an oscillator]{Charge transfer and coherence dynamics of tunnelling system coupled  to a harmonic oscillator }

\author{S Paganelli and S Ciuchi} 
\address{Dipartimento di Fisica, Universit\`{a} dell'Aquila,
Via Vetoio,I-67100 L'Aquila, Italy}
\address{CRS SMC, INFM-CNR, Roma Italy}
\ead{\mailto{simone.paganelli@roma1.infn.it},\mailto{sergio.ciuchi@aquila.infn.it}}

\date{\today}

\begin{abstract}

We study the transition probability and coherence of a two-site system,
interacting  with an oscillator. Both properties depend on the initial 
preparation. The oscillator is prepared in a thermal state and, even though it
cannot be considered as an extended  bath, it produces decoherence because of
the large number of states involved in the dynamics.  In the case in which the
oscillator is initially displaced a coherent dynamics of change entangled with
oscillator modes takes place. Coherency is however degraded as far as the
oscillator mass increases producing a increasingly large recoherence time.
Calculations are carried on by exact diagonalization and compared with two
semiclassical approximations. The role of the quantum effects are highlighted in
the long-time dynamics, where semiclassical approaches give rise to a 
dissipative behaviour. Moreover, we find that the oscillator dynamics has to be
taken into account, even in a semiclassical approximation, in order to
reproduce a thermally activated enhancement of the transition  probability.   

\end{abstract}

\submitto{\JPCM}


\section{Introduction}
\label{intro}

The quantum dynamics of a charge, moving between two potential minima,
is strongly influenced by the variations of the surrounding geometrical configuration.
The potential the charge is put in, is usually produced by  heavy degrees of freedom (a lattice, a molecular structure or an environment) evolving as well.
In many cases, only one normal mode of the heavy system is expected to be coupled with the tunnelling charge.
This occurs for very simple molecular or mesoscopic structure, or when the time-scales of heavy and 
light systems' dynamics are so different to allow the coupling with a single collective mode.      

As an example, we can consider the transport of a charge between localized sites, in a 
crystal, affected by coupling with optical
phonon modes, hereafter we shall refer to this picture also for the choice of the notation.
Moreover, there are a lot of other cases formally similar to it.    
Another example is the electron transfer reaction, in which an electron is excited 
toward an excited molecular  state which is strongly coupled to ionic motion. In
this case, coupling  can reduce the tunnelling frequency
of the electron between the two molecular states. The result is a freezing of the
electron into a definite excited state, in which electronic and associated ionic state are entangled 
\cite{jean:10464}.
As a third example, we may consider a single-molecular conductor  made by carbon nanotube
\cite{Leroy}. Here, the negative differential conductance behavior is 
associated with phonon-mediated electron tunnelling processes
\cite{FeinbergZazunovPRB06}.

In all these cases, the behaviour of the systems qualitatively  changes, 
as the temperature of the oscillator increases, giving rise to a 
charge transfer process which continuously changes from coherent  quantum
tunnelling to incoherent classical hopping.
Coherence properties of tunnelling systems are now accessible to a wide class of
experiments. By broadband absorption spectroscopy, which is able to access
time-resolved kinetics, it is possible to detect coherent oscillations in 
excited-state electron transfer of fluorinated benzenes\cite{kovalenko}.
In a single quantum dot, Rabi oscillations have been detected using quantum wave
function interferometry \cite{KamadaPRL01}. Here, the the electromagnetic field
strongly couples with excitonic levels of the dot.

Because of the large differences among all these physical cases,  
the corresponding tunnelling system lives in very different regions of
parameters space. For example,  in molecular systems, 
the oscillator  frequency can be much larger than the
tunnelling amplitude of the system, leading to an {\it antiadiabatic}
regime.  While, in solid state physics dispersionless oscillators can modelize
optical phonons of the systems whose frequencies are often much lower than hopping
amplitudes of the itinerant electrons, leading to adiabatic regime.

To perform an extensive study of coherence and tunnelling, as system's
parameters span all the accessible phase diagram, we choose  a simple model of a
single tunnelling system, coupled to a single oscillator. Such a model  has
been  widely analyzed, in different regimes and approximations, because of its
importance both as a building  block of cluster expansion in a lattice
model\cite{ammberciu}, and as a model for chemical reactions and charge transfer
in organics. Following a block diagonalization   technique, introduced by Fulton
and Gouterman \cite{FG}, we calculate quantum propagators. These results has
been used for calculating exactly  the finite temperature spectral functions
\cite{paganelciuk}   and to characterize the band-width behaviour with
temperature.  Here, we study the quantum dynamics of the system, considering the
transition probability and coherence. For this purpose, we study  the reduced
density matrix, taking into account two different initial system preparations: 
i) the {\it electron} preparation, in which oscillator is taken from an
thermal equilibrium distribution in absence of interaction, and  ii){\it polaron}
preparation, in which oscillator's initial state is taken from a thermal equilibrium
distribution in absence of tunnelling.  Notice that the two considered
preparations are also representative of systems in which the charge is promoted
to a given energy level with (polaron) or  without (electron) vibronic
relaxation. Reduction is obtained by tracing out the bosonic degrees  of
freedom. The particle transitions are characterized by the diagonal elements,
while the degree of coherence is given by the so-called purity which give 
a quantitative indication of how the system's state is a pure quantum state. 
The temporal behaviour of these quantities depends on the  coupling strenght, as
well as the adiabaticity parameter,  i.e. the ratio between tunnelling and
oscillator characteristic times. 

In the antiadiabatic regime, when the oscillator is faster than the tunnelling
system, polaron preparation guarantees a coherent behaviour up to temperatures
of the order of the phonon frequency. On the contrary,  the  electron
preparation gives rise to a fast decoherence because of the entanglement with
the oscillator mode,  which produces a dissipative effect even at zero
temperature. As temperature increases, both preparations gives the same
incoherent dynamics.



In the adiabatic regime, the phonon spectrum tends to a continuum and polaron
recoherence times becomes  longer and  longer. We observe a decoherence in both
electron and polaron preparation.  In this regime, we compare the exact result
with a  static (SA) and a Quantum-Classical (QC)  approximation yet introduced
in \cite{paganelciuk2}. We find that QC approximation is able to capture the
high temperature polaron incoherent motion, while  SA is sufficiently accurate
only for short times.

%

%
As a drawback of the finiteness of the system, the equilibrium is never really
reached and in principle  infinite recoherences appear. This purely quantum
behaviour is not recovered neither by  SA nor by  QC semiclassical
approximations. To observe a real dissipation it necessary to introduce a
reservoir  with a continuum spectral density. Nevertheless, we expect that in an
intermediate time-scale,  between the initial dynamics, driven by the fast
degree of freedom,  and  the recoherence times, our single-oscillator model
reproduces the many-oscillator case, providing that a  mode with dominant
interaction can be separated from the rest of the bath.


The model is described in Sec. \ref{sec:model}. In Sec. \ref{sec:redmat}
we introduce the reduced density  matrix 
for the polaron and the electron. In Sec. \ref{sec:analytics}
we describe the exact mapping, by means of Fulton-Gouterman transformations,
from the original electron-phonon problem into two single anharmonic oscillators.
Then the QC and SA approximation are described.
In Sec. \ref{sec:results} are presented the results and the 
comparison between these three different techniques is discussed. Sec. \ref{sec:concl} is
devoted to the conclusion.

\section{The Model}
\label{sec:model}

The model we shall consider is described by the following Hamiltonian
\begin{equation}\label{eqn:holstein}
  H = \omega_0 a^\dag a-J \sigma_x-\tilde{g}\sigma_z(a^\dag+a),
\end{equation}
describing a spin-$1/2$ interacting with a harmonic oscillator of frequency $\omega_0$.
The model can be associated to a large number of 
physical systems \cite{swain-cfracsol,cini-recmethod} 
but, for sake of clearness, we shall refer to an electron,  in the tight
binding approximation, moving in a two-site lattice and interacting with
it by the local distortion of the lattice site \cite{paganelciuk}.
In particular, it can be shown that it is equivalent to the Holstein two-site model
\cite{holstein,firsov,paganelciuk}, with  operators $a$ and $a^\dag$  
referring to the relative  phonon coordinate
and providing  that fermionic operators  are mapped into 
a pseudo-spin notation $\sigma_z = c_1^\dag c_1 - c_2^\dag c_2 $
and $\sigma_x = c_1^\dag c_2 + c_2^\dag c_1$.
The center of mass coordinate   can easily be decoupled
(for a more detailed discussion see \cite{paganelciuk, ranninger}).
Therefore throughout this paper we shall refer the word electron to the tunneling system 
and the word phonon to the oscillator.

The strength of the electron-phonon interaction is given by the constant
$\tilde{g}=g / \sqrt{2}$,  $J$ is the electron wave function overlap or hopping
and $2J$ is the tight binding half-bandwidth.

Beside the temperature, we can choose two parameters that characterize the
model  i) the bare e-ph coupling constant $\lambda=g^2/(\omega_0 J)$ given by
the ratio of the   polaron energy ($E_p=-g^2/\omega_0$) to the hopping
$J$ and ii) the adiabatic ratio $\gamma = \omega_0/J$. 

In terms of these parameters we can define a weak-coupling $\lambda<1$ and
strong coupling $\lambda>1$ regimes, as well as an adiabatic $\gamma<1$ or
anti-adiabatic $\gamma>1$ regimes.

Notice that instead of choosing $\lambda$ as coupling constant we may choose
another combination which is more appropriate in the so called atomic ($J=0$)
limit i.e. $\alpha=\sqrt{\lambda/(2 \gamma)}$ (see Appendix).

\section{Reduced Density matrix}
\label{sec:redmat}

The study of the charge dynamics is not trivial because, 
in general,  it is entangled with the 
harmonic oscillator. 
The time dependent correlation functions of the two-site 
Holstein model  has been investigated in the past \cite{demello} 
and also a short time transfer dynamics has been introduced in \cite{herfortfilosofico}.

In this paper, we  introduce a density matrix approach for the charge dynamics over a 
very large time range. 
Hereafter, we shall assume that charge and oscillator 
are initially separated, being the former localized on 
the first site and the latter in a mixed thermal state.
The corresponding density matrix is
\begin{equation}
\rho(0)=\sum_n \frac{e^{-\beta \omega_0 n}}{Z}
\proj{\phi_n}{\phi_n}\otimes\proj{1}{1},
\end{equation}
where we used the notation $\ket{1}=c_1^\dagger \ket{0}$ and $\beta$ is the
inverse temperature. The state
$\ket{\phi_n}$  depends on the choice of the initial preparation 
\cite{lucke}, in this paper we  study two different situations obtained from two
different limiting regimes:

\begin{enumerate}
\item \emph{electronic preparation} \EP:  
the electron is initially free ($g=0$) and
the oscillator is at its thermal equilibrium 
\begin{equation}
\rho^{(el)}(0)=\sum_n \frac{e^{-\beta \omega_0 n}}{Z}
\proj{n}{n}\otimes\proj{1}{1},
\label{eqn:elprep}
\end{equation}

\item \emph{polaronic preparation} \PP:  
electron is initially localized ($J=0$) on a given site (say $1$),  
while the oscillator is displaced accordingly (see Appendix) 
\begin{equation}
\rho^{(pol)}(0)=\sum_n \frac{e^{-\beta \omega_0 n}}{Z}
\proj{\psi^1_n}{\psi^1_n}\otimes\proj{1}{1},
\label{eqn:polprep}
\end{equation}

\end{enumerate}
The dynamics is obtained by switching on  $g$, in the first case, and $J$, in
the second one, and letting evolve the density matrix with  the Hamiltonian  
(\ref{eqn:holstein}) $\rho(t)=e^{-i Ht} \rho(0) e^{i H t}$. The temperature
enters {\it only} in the initial state  trough the incoherent distribution of
the initial oscillator states in both preparation.

Tracing over the oscillator degree of freedom, we obtain the electron reduced density matrix
\begin{equation}
\label{rhoel} 
    \rho^{(el)}(t)=\Tr_{ph}\{\rho(t)\},
\end{equation}
which, in terms of the oscillator's number states, is
\begin{equation} 
\label{rhoel1}
\rho^{(el)}(t) =\sum_{n,m} \frac{e^{-\beta \omega_0 n}}{Z}
    \bra{m}e^{-i  {H} t}\proj{n,1}{n,1}e^{i
    {H}t}\ket{m},
\end{equation}

To  characterize the motion of the polaron we cannot  reduce the
density matrix  by  tracing out the phonon degrees of
freedom, this is because the polaron itself contains phonons.
In order to understand better the polaron dynamics, let us first 
apply a Lang-Firsov transformation $D$ (see Appendix),
the new fermionic particle corresponds to a polaron, so the 
density matrix  with the initial localized polaron can be written as
\begin{equation}\label{eqn:poldensm}
\label{rhopol} 
\rho^{(pol)}(t) = Tr_{ph}\{D^\dag\rho(t)D \},
\end{equation}
and reads, in terms of the oscillator's number states, as
\begin{equation}
\label{rhopol1}
\rho^{(pol)}(t) =\sum_{n,m} \frac{e^{-\beta \omega_0 n}}{Z}
    \bra{m}e^{-i \bar{H} t}\proj{n,1}{n,1}e^{i
   \bar{H}t}\ket{m}.
\end{equation}

\subsection{Quantities of interest}
In this paper, we will study two measures:
one for transition probability and the other for the degree of coherence. 
The diagonal elements of the reduced density matrix, in the site basis,
represent the population of each site. The
transition probability from site 1 to site 2 is given by
\begin{equation}
\label{trans}
w_{1,2}(t)=\bra{2}\rho(t)\ket{2}.
\end{equation}
where ${\rho}$ is the {\it reduced} density matrix
in any of the previously introduced preparations.
 
The off-diagonal elements of the reduced density matrix 
represent the quantum interference  between localized amplitudes. However their
knowledge is not sufficient to determine whether the state is pure or not.
Suppose that initial state is pure, if diagonal elements do not evolve in time,
the suppression of the off-diagonal elements implies the evolution into a mixed
state. In this particular case, the knowledge of off-diagonal elements also
determines the purity of the system. In the more general case in which all the
elements of $\rho$ evolve, the choice of the off-diagonal elements obviously
depends on the basis. A  basis independent measure for purity (called
\emph{purity} itself) is 
\begin{equation}
\label{purity}
P(t)=\Tr{{\rho}^2(t)}.
\end{equation}
where again ${\rho}$ is the reduced density matrix. It is easy to see that $1/2
\leq P \leq 1$  with  $P=1$ if and only if the state is pure and $P=1/2$ when
the state is maximally mixed. 

The behaviour of our finite system results as a superposition of oscillations
with many different characteristic frequencies. To disentangle the relevant
timescales at a given time $t$ it is found useful to consider the time averaged
transition probability and coherence, defined as
\begin{equation}
\label{timeav}
\bar{Q}(t)=\frac{1}{t}\int_0^t dt' Q(t'),
\end{equation}
where $Q$ can be either $w_{1,2}$ or $P$.

\section{Methods}\label{sec:methods}
\label{sec:analytics}

In this section we present the methods which we use to get 
the reduced density matrices for both initial preparations.

\subsection{Exact diagonalization}

As shown by Fulton and Gouterman \cite{FG}, a two-level system coupled to an
oscillator in such a manner that the total Hamiltonian displays a
reflection symmetry, may be subjected to a unitary transformation which
diagonalizes the system with respect to the two-level subsystem
\cite{FG,wagner,wagner3,feinberg90}.
This method can be generalized to the N-site situation, if the symmetry of
the system is governed by an Abelian group \cite{wagner3}.

In particular, an analytic method for calculating the Green functions of the two-
site Holstein model is
given in \cite{paganelciuk,ciuchi2}. Here, the Hamiltonian is diagonalized in the 
fermion
subspace by applying a Fulton Gouterman (FG) transformation. So the initial problem 
is  mapped into
an effective  anharmonic oscillator model. It is possible to introduce  different 
FG transformations
for the electron and the polaron. The new problem results to be very simplified and 
very  suitable to be numerically implemented. Analytical continued-fraction results 
exist for the electron case\cite{paganelciuk,ciuchi2}.

In this section, we briefly recall the FG transformations method. 
The density matrix elements are given explicitly in terms of effective Hamiltonians and calculated
by means of exact diagonalization

The FG transformation  we use for the electronic case is 
\begin{equation}\label{eqn:can}
  V = \frac{1}{\sqrt{2}}\left(%
\begin{array}{cc}
  1  & (-1)^{a^\dag a}\\
  -1 & (-1)^{a^\dag
a}\\
\end{array}
\right),
\end{equation}
the new Hamiltonian $\tilde{H}=V H V^{-1}$ becomes diagonal in the 
electron subspace
\begin{equation}\label{eqn:acca}
    \tilde{H}=\left(
    \begin{array}{cc}
  H_+  & 0\\
  0 & H_- \\
\end{array}\right),
\end{equation}
the diagonal elements, corresponding  to the bonding and antibonding 
sectors of the electron subspace, being two purely phononic Hamiltonians 
\begin{equation}\label{eqn:hamph}
    H_\pm=\omega_0 a^\dag a \mp J(-1)^{a^\dag a}-\tilde{g}(a^\dag+a).
\end{equation}
The operator $(-1)^{a^\dag a}$ is the reflection operator in the vibrational 
subspace and it satisfies the condition
$(-1)^{a^\dag a}a(-1)^{a^\dag a}=-a$.
A wide study of the eigenvalue problem was carried out
in \cite{herfort1}, both numerically and analytically, by a variational method, 
extending the former results given in \cite{ranninger}.
In \cite{herfort1} $  H_\pm $ is approximately diagonalized by applying 
a displacement, the dynamics is reconstructed by 
the calculated eigenvectors and energies. 

The evaluation of the polaron Green function can be done on the same footings,
but the expression involves also the non diagonal elements of the resolvent
operators, causing  an exponential increasing of the numerical calculations.

To avoid this problem, we first perform  the LF transformation   
and then apply, on the resulting Hamiltonian (\ref{eqn:parappa}),
a different FG transformation 
\begin{eqnarray}\label{eqn:can1}
  V_1 = \frac{1}{\sqrt{2}}\left(
\begin{array}{cc}
  1             & -(-1)^{a^\dag a}\\
  (-1)^{a^\dag a} &   1\\
\end{array}
\right).
\end{eqnarray}
The new Hamiltonian $\tilde{H}_{LF}=V_1 \bar{H} V_1^{-1}$ is
\begin{equation}\label{eqn:acca1}
    \bar{H}_{LF}=\left(
    \begin{array}{cc}
  \bar{H}_+  & 0\\
  0 & \bar{H}_- \\
\end{array}\right),
\end{equation}
where
\begin{equation}\label{eqn:hbarra}
\bar{H}_\pm =\omega_0 a^\dag a+J(-1)^{a^\dag a} e^{\mp 2\alpha (a^\dag-a)}+E_p/2,
\end{equation}
is real and symmetric but not tridiagonal in the basis of the harmonic
oscillator, the matrix elements of $\bar{H}_\pm$ are given in \cite{paganelciuk}. 

In order to write down the density matrix elements, let us introduce the following notation:
\begin{eqnarray}\label{eqn:risolventiel}
    R_{m,n}^{(\pm)}(t) &=&\bra{m}e^{-i\ham_\pm t}\ket{n}\\
    \bar{R}_{m,n}^{(\pm)}(t)&=&\bra{m}e^{-i\bar{\ham}_\pm t}\ket{n},
\end{eqnarray}
\begin{eqnarray}
N_{1,1}^{m,n}(t)&=&\bra{m,1}e^{-i\ham t}\ket{n,1}=
    \frac{1}{2}\left[ R_{m,n}^{(+)}(t)+R_{m,n}^{(-)}(t)\right]\\
N_{2,1}^{m,n}(t)&=&\bra{m,2}e^{-i\ham t}\ket{n,1}=
    \frac{(-1)^m}{2}\left[ R_{m,n}^{(+)}(t)-R_{m,n}^{(-)}(t)\right],
\end{eqnarray}
\begin{eqnarray}
M_{1,1}^{m,n}(t)&=&\bra{\psi_m^{1},1}e^{-i\ham t}\ket{\psi_n^{1},1}
=\frac{1}{2}\left[\bar{R}_{m,n}^{(+)}(t)+(-1)^{m+n} \bar{R}_{m,n}^{(-)}(t)\right] \\
M_{2,1}^{m,n}(t)&=&\bra{\psi_m^{2},2}e^{-i\ham t}\ket{\psi_n^{1},1}=\frac{1}{2}\left[(-1)^{n}\bar{R}_{m,n}^{(-)}(t)-(-1)^{m} \bar{R}_{m,n}^{(+)}(t)\right].
\end{eqnarray}

The reduced electron density matrix elements are
\begin{eqnarray}
    \rho^{(el)}_{1,1}(t)&=& \sum_{n,m} \frac{e^{-\beta \omega_0 n}}{Z} |N_{1,1}^{m,n}(t)|^2\nonumber 	\\
	\rho^{(el)}_{2,1}(t)&=&\sum_{n,m} \frac{e^{-\beta \omega_0 n}}{Z}
     N_{2,1}^{m,n}(t) N^{ * m,n}_{1,1}(t), 
     \label{eqn:rhoelFG}
\end{eqnarray}
the calculation for the polaron  case gives
\begin{eqnarray}
    \rho^{(pol)}_{1,1}(t)&=&
    \sum_{n,m} \frac{e^{-\beta \omega_0 n}}{Z}
    |M_{1,1}^{m,n}(t)|^2 \nonumber \\
    \rho^{(pol)}_{2,1}(t)&=&
    \sum_{n,m} \frac{e^{-\beta \omega_0 n}}{Z}
     M_{2,1}^{m,n}(t) M^{ * m,n}_{1,1}(t).
     \label{eqn:rhopolFG}
\end{eqnarray}

A qualitative insight into the relevant timescales involved in the evolutions of
$ \rho^{(el)}$ and $ \rho^{(pol)}$ can be gained by looking at the behaviour of
the spectral functions of the model (\ref{eqn:holstein}) studied in our
previous work \cite{paganelciuk}. In terms of the Fourier transform of function 
$N_{1,1}^{m,n}(t)$, the electron spectral function $A(\omega)$ 
can be defined as
\begin{equation} A(\omega)= -\frac{1}{\pi} 
Im \sum_{n} \frac{e^{-\beta \omega_0 n}}{Z} N_{1,1}^{n,n}(\omega). 
\label{eqn:spectralrho} 
\end{equation} 
Analogous equation holds for the polaron spectral function relating it to the
function $M_{1,1}(\omega)$.

\begin{figure}[htbp]
\begin{center}
\includegraphics[width=\figW,angle=-0]{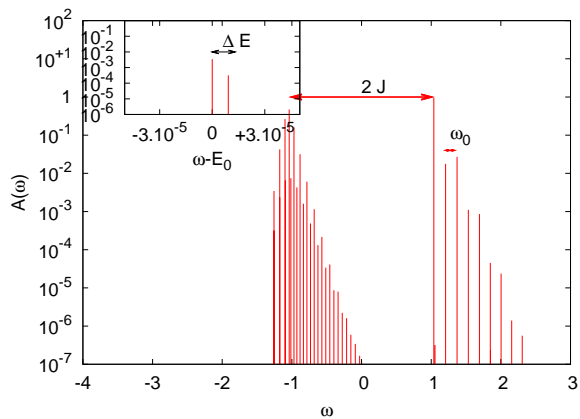}
\end{center}
\caption{Electron spectral function for $\gamma=0.1$ $\lambda=2$and $T=0$ 
see ref. \cite{paganelciuk}.}
\label{fig:spectral}
\end{figure}

An example of $A(\omega)$ is reported in figure \ref{fig:spectral}.  
We notice that three energy scales (depicted schematically in figure
\ref{fig:spectral})  can be associated  $A(\omega)$ \cite{paganelciuk}.
One is the separation of the low lying energy level $\Delta E$,
the other is the phonon energy $\omega_0$ and finally there is the tunnelling $J$.
They are depicted schematically in figure \ref{fig:spectral}. 
These energy scales define three different timescales:
 
i) $\tau_J=2\pi J^{-1}$,

ii) $\tau_{\omega_0}=2\pi\omega_0^{-1}$ 

iii) $\tau_Q=2\pi\Delta E^{-1}$.

As it is reasonable from the relation between spectral functions and  reduced
density matrix (\ref{eqn:spectralrho}), these characteristic
timescales are recovered in the reduced density matrix evolution.

\subsection{The static approximation}\label{sub:adiab}
The case, in which a light quantum particle interacts with  much more massive particles,  is
very common in solid state and molecular physics.
We discuss the \emph{adiabatic regime}, meaning that, 
in a characteristic time for the  the light 
particle dynamics, the heavy degrees of freedom can be 
considered approximately quiet. Here, we describe the SA 
approach in its basic formulation for the dynamics.

The Hamiltonian (\ref{eqn:holstein}) can be written 
in the coordinate-momentum representation 
\begin{equation} \label{eqn:Hadiab}
    \ham=\frac{p^2}{2 m}+\frac{m
\omega_0^2}{2}x^2-\frac{\bar{g}}{\sqrt{2}}x\sigma_z-J\sigma_x-\omega_0,
\end{equation}
with $\bar{g}=g \sqrt{2m \omega_0}$.
In the adiabatic limit ($\gamma\ll1$), the phonon 
is much slower than the electron
(heavy phonon and large electron tunnelling amplitude) and one can neglect the phonon kinetic
term in (\ref{eqn:Hadiab}). This is the well known Born-Oppenheimer approximation. In practice, 
 it consists in studying  the electronic problem with $x$ as a classical parameter.

Within this approximation, we put $\omega_0=\sqrt{k/m}\rightarrow 0$ ($m$ is the ion mass) and  the Hamiltonian becomes
\begin{equation}\label{eqn:relham}
    \ham_{ad}=\frac{k}{2}x^2-\frac{\bar{g}}{\sqrt{2}}x \sigma_z-J\sigma_x.
\end{equation}
The eigenvalues can be expressed trough the classical displacement $x$
\begin{equation}
\label{eqn:adiabatic_energies}
V_\pm(x)=\frac{k}{2}x^2 \pm \Omega(x),
\end{equation}
with $\Omega(x)=\sqrt{\frac{\bar{g}^2}{2}x^2+J^2}$.
The lowest branch ($-$) of  (\ref{eqn:adiabatic_energies})
defines an adiabatic potential which has a minimum at $x=0$ as far as
$\lambda<1$ while for $\lambda>1$, it becomes double well potential with minima at 
$\pm x_m$, $x_m=\sqrt{\frac{\bar{g}^2}{2 k^2}-\frac{2 J^2}{\bar{g}^2}}$, in this case  the electron is mostly localized on a
given site. The  quantum fluctuations are able to restore the symmetry
in analogy to what happens for an infinite lattice \cite{Lowen}. It is worth 
noticing that, in this limit, Hamiltonian  (\ref{eqn:holstein}) is equivalent to
the adiabatic version of the spin-boson Hamiltonian \cite{weiss,petruccione}.

The temporal evolution  is given by
\begin{equation}\label{eqn:evoluzio}
    e^{-i \ham_{ad} t}=e^{-i \frac{k x^2}{2} t}[\cos \Omega(x) t +i
    (\frac{\bar{g}x}{\sqrt{2}\Omega(x)}\sigma_z+\frac{J}{\Omega(x)}\sigma_x)\sin 
\Omega(x) t],
\end{equation}
so the density matrix dynamics can be explicitly calculated

The electronic initial preparation, corresponds to the density matrix 
\begin{equation}
\rho(0)=\proj{1}{1}\sqrt{\frac{k\beta }{2 \pi}} \int dx \,
e^{-\frac{\beta k}{2} x^2}\proj{x}{x},
\end{equation}
tracing out the phonon we obtain the electron reduced density matrix with elements
\begin{eqnarray}
\rho^{(el)}_{2,2}&=&\sqrt{\frac{\beta J \lambda}{2 \pi}}
\int du \, e^{-\frac{\beta J \lambda}{2} u^2}
\frac{\sin^2(J t \sqrt{u^2 \lambda^2+1)}}{1+\lambda^2 u^2}\nonumber\\
\rho^{(el)}_{1,2}&=&-i\sqrt{\frac{\beta J \lambda}{2 \pi}}
\int du \, e^{-\frac{\beta J \lambda}{2} u^2}
\frac{ \sin(2 J t \sqrt{u^2 \lambda^2+1}) }{2\sqrt{\lambda^2 u^2+1}}
\end{eqnarray}
where the scaled lenght $u=x k \sqrt{2}/\bar{g}$ was introduced.

In the same way, we can introduce the polaronic preparation
\begin{equation}
\rho(0)=\proj{1}{1}e^{-\frac{\beta \bar{g}^2}{4k}    }\sqrt{\frac{k\beta }{2 \pi}} \int dx
e^{-\beta(
    \frac{k}{2}x^2-\frac{\bar{g}}{\sqrt{2}}x)} \proj{x}{x},
\end{equation}
It is worth noting that, in the adiabatic limit, we cannot define the polaronic dynamics, as 
introduced in (\ref{eqn:poldensm}), because the operator $D$ is not defined for $\omega_0=0$.
In this limit, we study electronic dynamics with an initial polaronic preparation.
The corresponding reduced density matrix is

\begin{eqnarray}
\rho^{(pol)}_{2,2}&=&\sqrt{\frac{\beta J \lambda}{2\pi}}
    \int_{-\infty}^\infty du\,e^{-\frac{\beta J
    \lambda}{2}(u-1)^2}
  \frac{\sin^2(J  t \sqrt{(u \lambda)^2+1} ) }{ (u \lambda)^2+1}
\nonumber\\
\rho^{(pol)}_{1,2}&=&\sqrt{\frac{\beta J \lambda}{2\pi}}
    \int_{-\infty}^\infty du\, e^{-\frac{\beta J\lambda}{2}(u-1)^2}
     \left[ \frac{u \lambda\sin^2 (J t \sqrt{(\lambda u)^2+1}) }{((\lambda u)^2+1)}
       \right.\nonumber\\
    &-&\left.i \frac{\sin(2 J t \sqrt{(\lambda u)^2+1})}
    {2 \sqrt{(\lambda u)^2+1}}  \right]
\end{eqnarray}

It is possible to show that $\rho^{(pol)}_{2,2}$  is actually the adiabatic limit   
of the diagonal element of the reduced  polaronic density matrix, while this is not true for the 
off-diagonal elements.

We want to stress that, in the SA approach, the phonon is completely \emph{static} because its momentum $p$ has been  neglected. Here, only the
initial phonon distribution plays a role, but during electron hopping, oscillator is taken to be fixed.

\subsection{A quantum-classical dynamics approximation}
\label{app:QC}
To account for dynamics of the slow variable,
a mixed quantum-classical dynamics can be introduced.
In the past, several schemes for quantum-classical 
dynamics has been proposed, for example starting from the 
Born-Oppenheimer (SA) adiabatic approximation for the ground 
state at each step and using a density functional 
Hamiltonian \cite{car-parrinello,selloni}.
Another approach, good for a  short time dynamics, consists 
in a mapping from the Heisenberg equations
to a classical evolution by an average over the initial 
condition\cite{golosov,stock:1561}.
Some schemes are based on the evolution of the density matrix coupled to 
a classical bath \cite{berendsen,morozov}. A systematic expansion over the mass ratio has also 
been done, starting from partial Wigner transform of 
the Liouville operator, in  \cite{ciccotti1999,ciccotti2001JChPh,sergi2006JChPh}.
The QC approximation we use is essentially that of refs. \cite{berendsen,morozov}.

Let us consider Hamiltonian (\ref{eqn:Hadiab}), where $x$ and $p$  
are assumed to be classical variables  which can be represented as the components of  a vector
$\mathbf{u}$. Then a QC state vector can be introduced as 
\begin{equation}
\mathbf{v}=\mathbf{u}\otimes \mathbf{\sigma}=\left(
\begin{array}{c}
x\\
p\\
\sigma_x\\
\sigma_y\\
\sigma_z
\end{array}
\right).
\end{equation}
The classical variables evolve with the Ehrenfest equations
\begin{equation}
\left\{
\begin{array}{ccl}
\dot{x}&=&\frac{p}{m}\\
\dot{p}&=&\frac{m\omega_0^2}{2}x-\frac{\bar{g}}{\sqrt{2}}\left\langle \sigma_z\right\rangle ,
\end{array}
\right.
\end{equation}
while the quantum variables evolves in the Heisenberg picture 
\begin{equation}
\left\{
\begin{array}{ccl}
\dot{\sigma}_x &=& -\sqrt{2} \bar{g} x \sigma_y        			\\
\dot{\sigma}_y &=&  \sqrt{2} \bar{g} x \sigma_x  -2J \sigma_z	\\ 
\dot{\sigma}_z &=& 	                              2J \sigma_y
\end{array}
\right..
\end{equation}
To give a unified description of the overall evolution, we define a Liouvillian 
operator $\mathcal{L}=\mathcal{L}_x+\mathcal{L}_p+\mathcal{L}_\sigma$
with 
\begin{equation}
\mathcal{L}_\sigma=-i \left(
\begin{array}{ccc}
 0 									& -\sqrt{2} \bar{g} x  	& 0 						\\
\sqrt{2} \bar{g} x	& 		0									& -2J 					\\
0										& 			2J 							&   0  
\end{array}
\right)
\end{equation}
and $\mathcal{L}_x = \dot{x} \frac{\partial}{\partial x} $ 
$\mathcal{L}_p = \dot{p} \frac{\partial}{\partial p}$.
So, the time evolution is given by  
\begin{equation}
\mathbf{v}(t)=e^{i \mathcal{L}t}\mathbf{v}(0).
\end{equation}

The numerical integration can be implemented using the symmetries Trotter breakup formula \cite{wan:4447,deraedt-trotter}
\begin{equation}
\mathbf{v}(t)\simeq \left(
e^{i \mathcal{L}_\sigma\frac{\epsilon}{2}}
e^{i \mathcal{L}_p     \frac{\epsilon}{2}}
e^{i \mathcal{L}_x           \epsilon}
e^{i \mathcal{L}_p     \frac{\epsilon}{2}}
e^{i \mathcal{L}_\sigma\frac{\epsilon}{2}}
\right)^N \mathbf{v}(0)
\end{equation}
with $\epsilon=t/N$. 
All the density matrix elements, can be expressed in terms of elements of $\mathbf{v}(t)$.


\section{Results}\label{sec:results}
\subsection{Antiadiabatic regime} 

\begin{figure}[htbp]
\begin{center}
\includegraphics[width=\figWc,angle=-0]{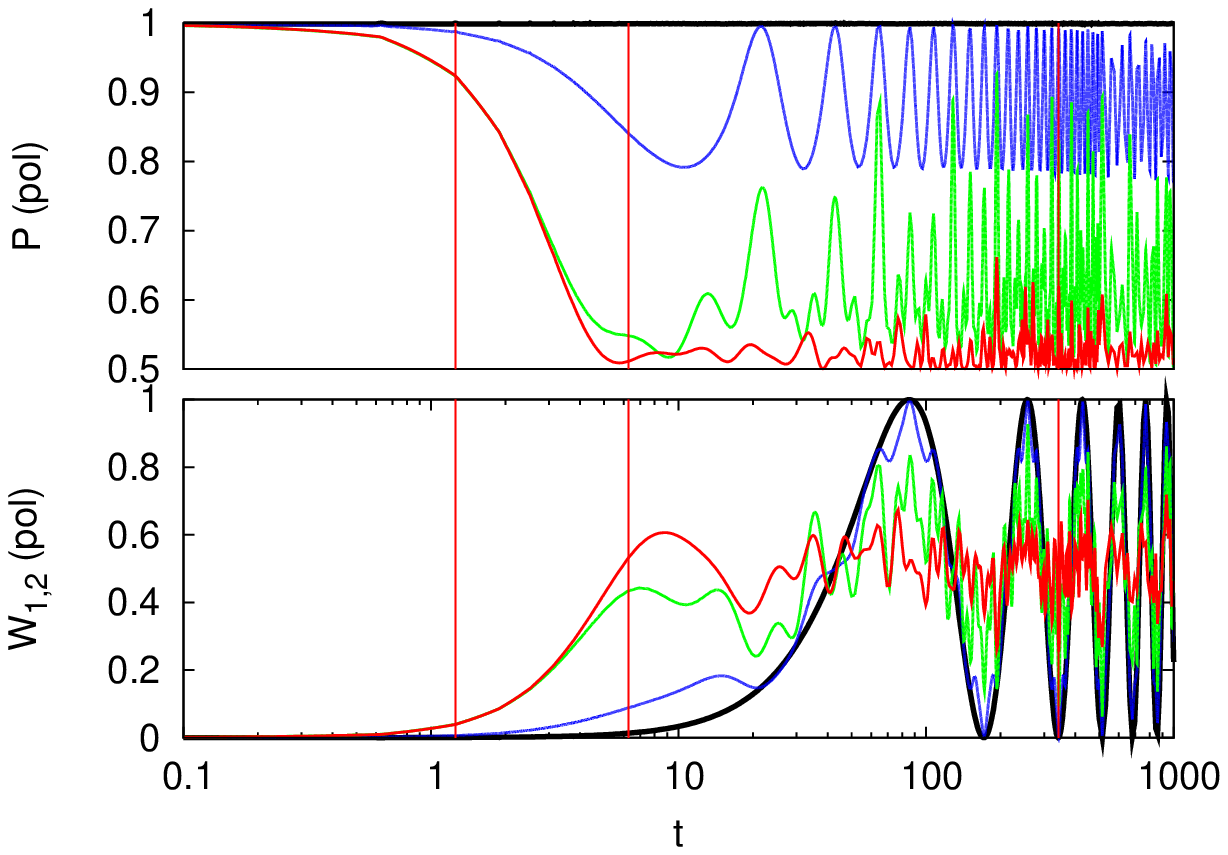}
\includegraphics[width=\figWc,angle=-0]{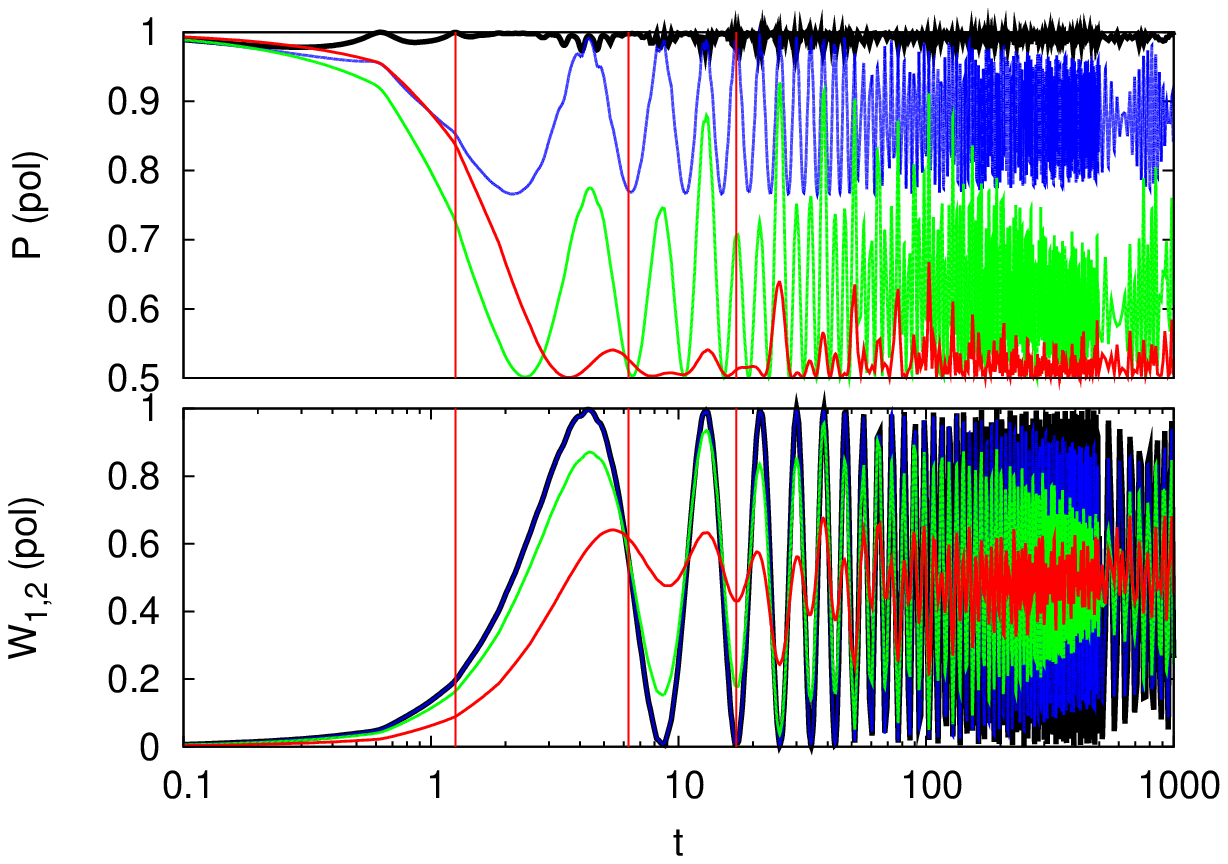}
\end{center}
\begin{center}
\includegraphics[width=\figWc,angle=-0]{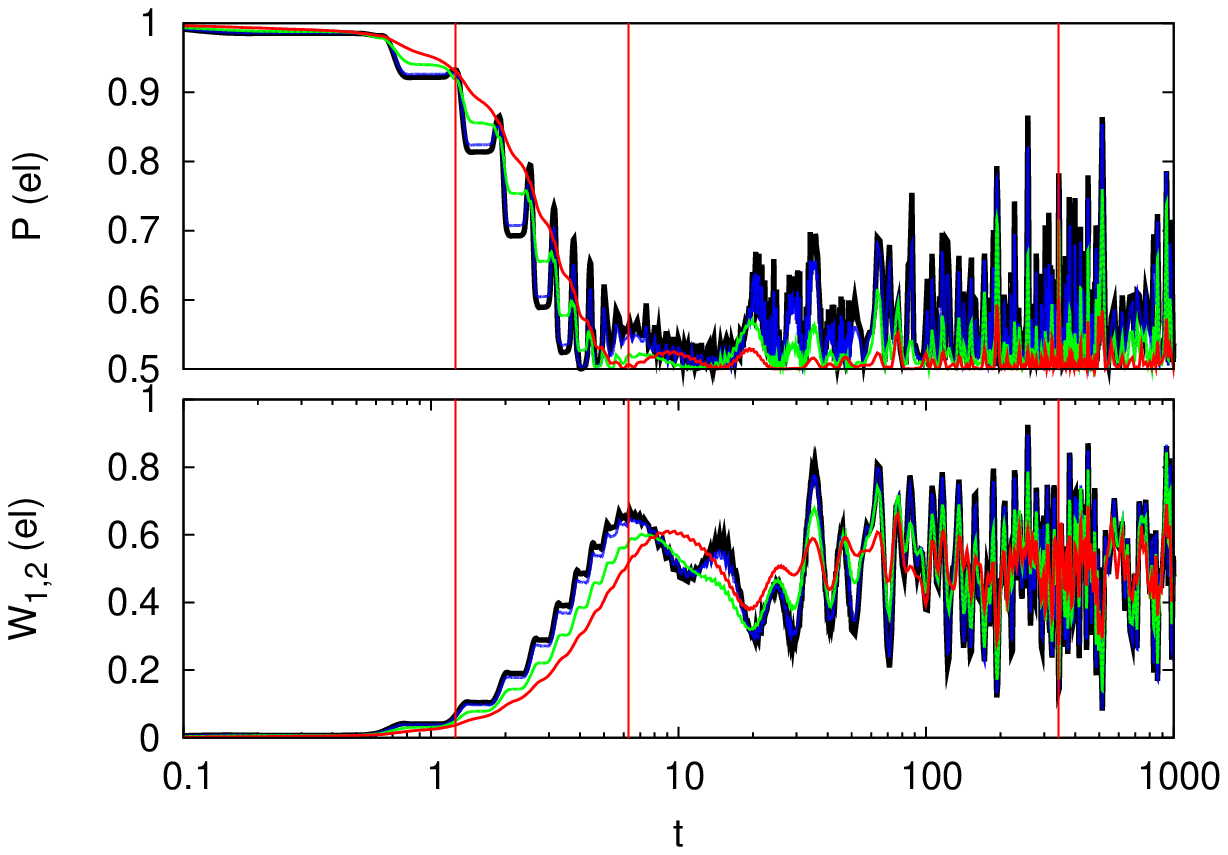}
\includegraphics[width=\figWc,angle=-0]{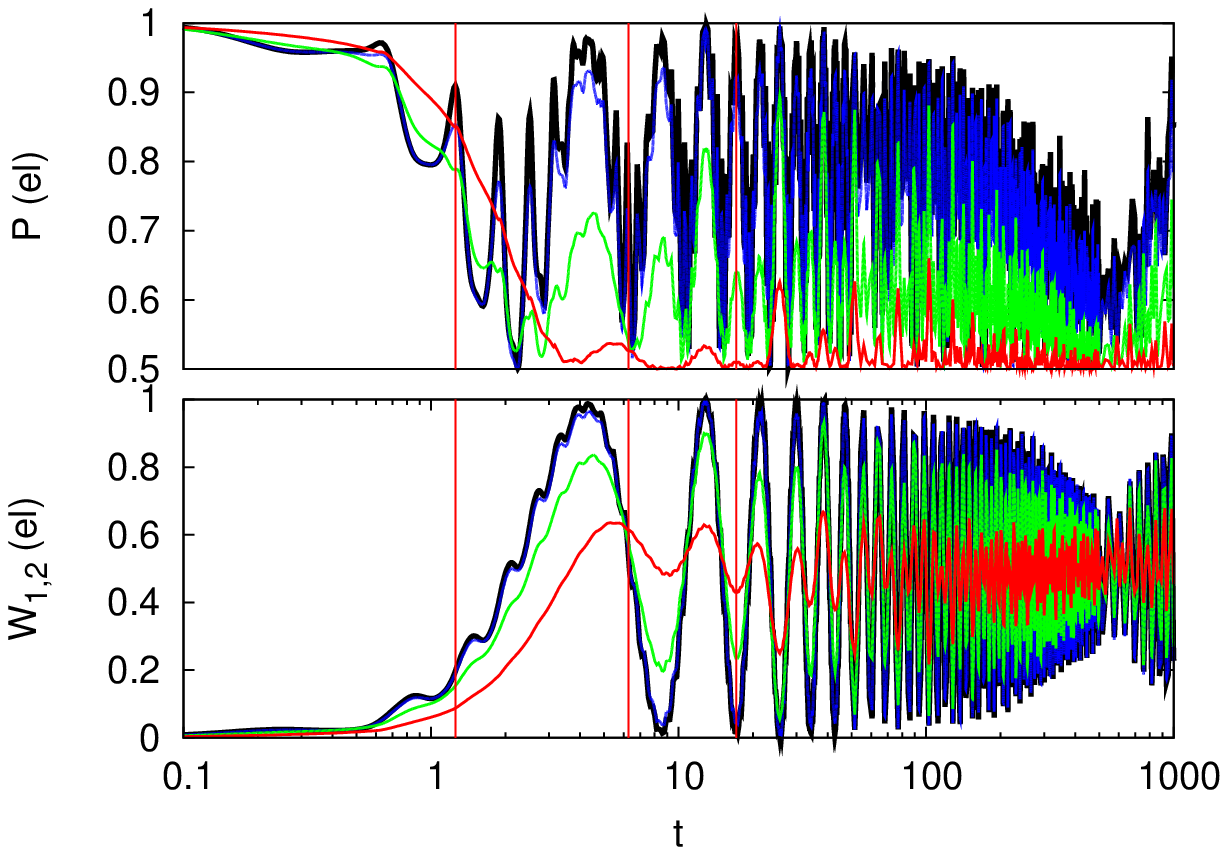}
\end{center}
\caption{Polaron (top) and electron (bottom) populations and purity. Left panels 
antiadiabatic and strong coupling regime: $\gamma=10$ $\lambda=40$.
Right panels antiadiabatic and weak coupling regime $\gamma=10$ $\lambda=10$.
Curves are for $T/\omega_0=0.0$ (black), $T/\omega_0=0.5$ (blue),
$T/\omega_0=2.0$ (green), $T/\omega_0=10.0$ (red). Vertical lines marks from
left to right the timescales $\tau_{\omega_0}$,$\tau_J$,$\tau_Q$.}
\label{fig:anti}
\end{figure}

\begin{figure}[htbp]
\begin{center}
\includegraphics[width=\figWc,angle=-0]{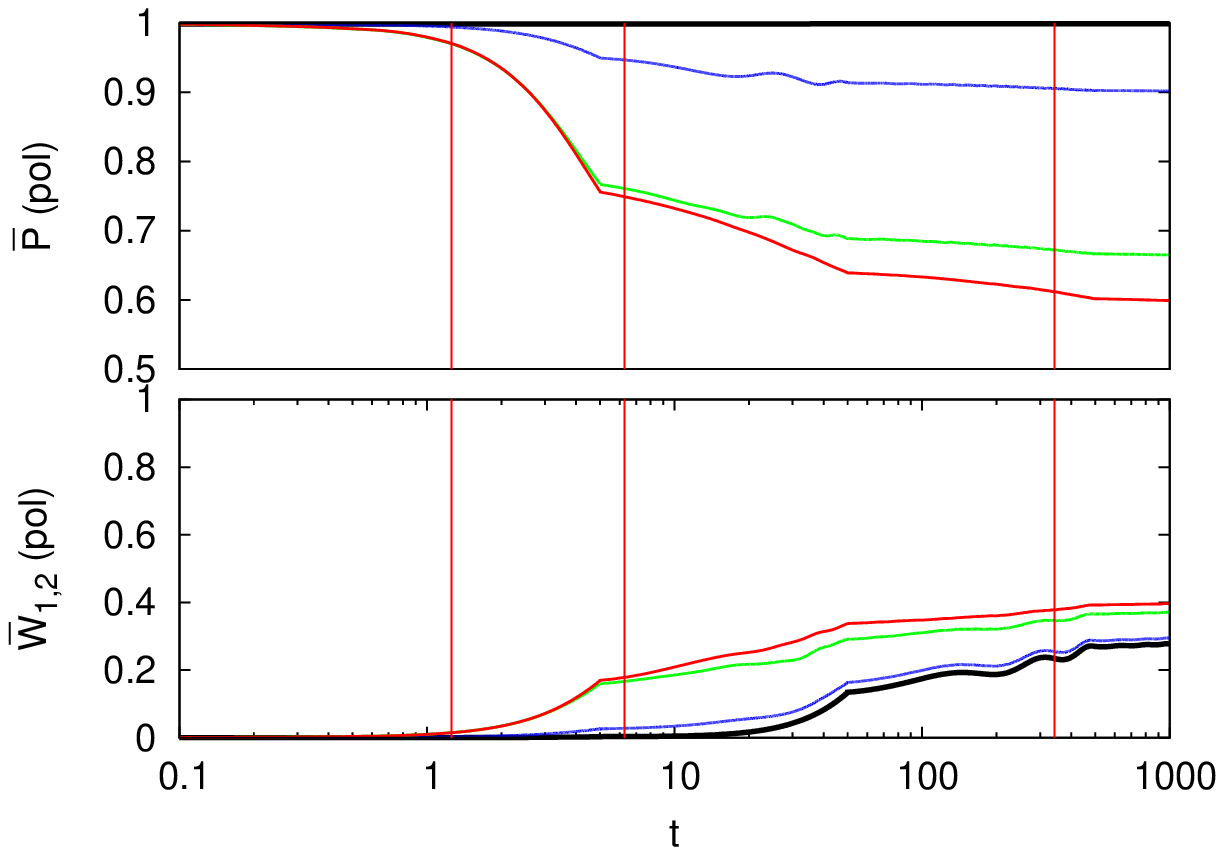}
\includegraphics[width=\figWc,angle=-0]{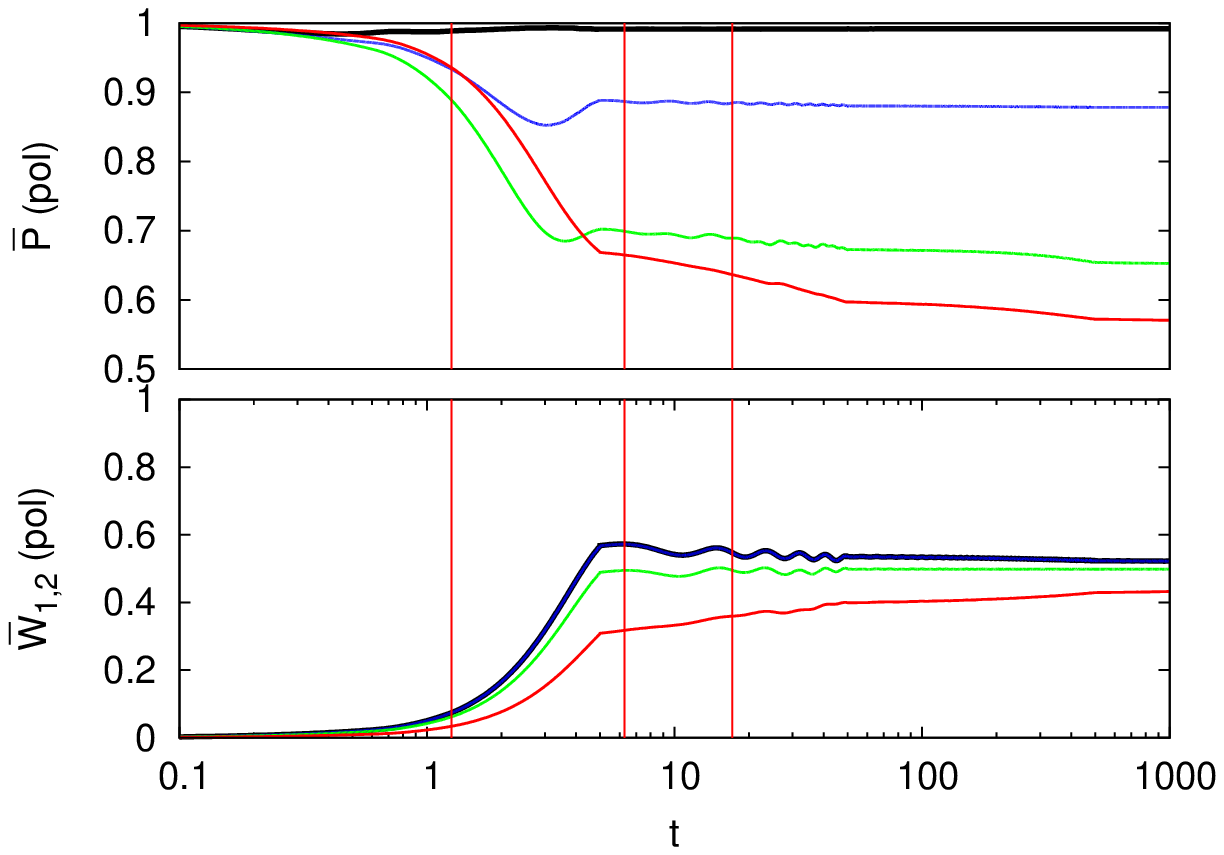}
\end{center}
\begin{center}
\includegraphics[width=\figWc,angle=-0]{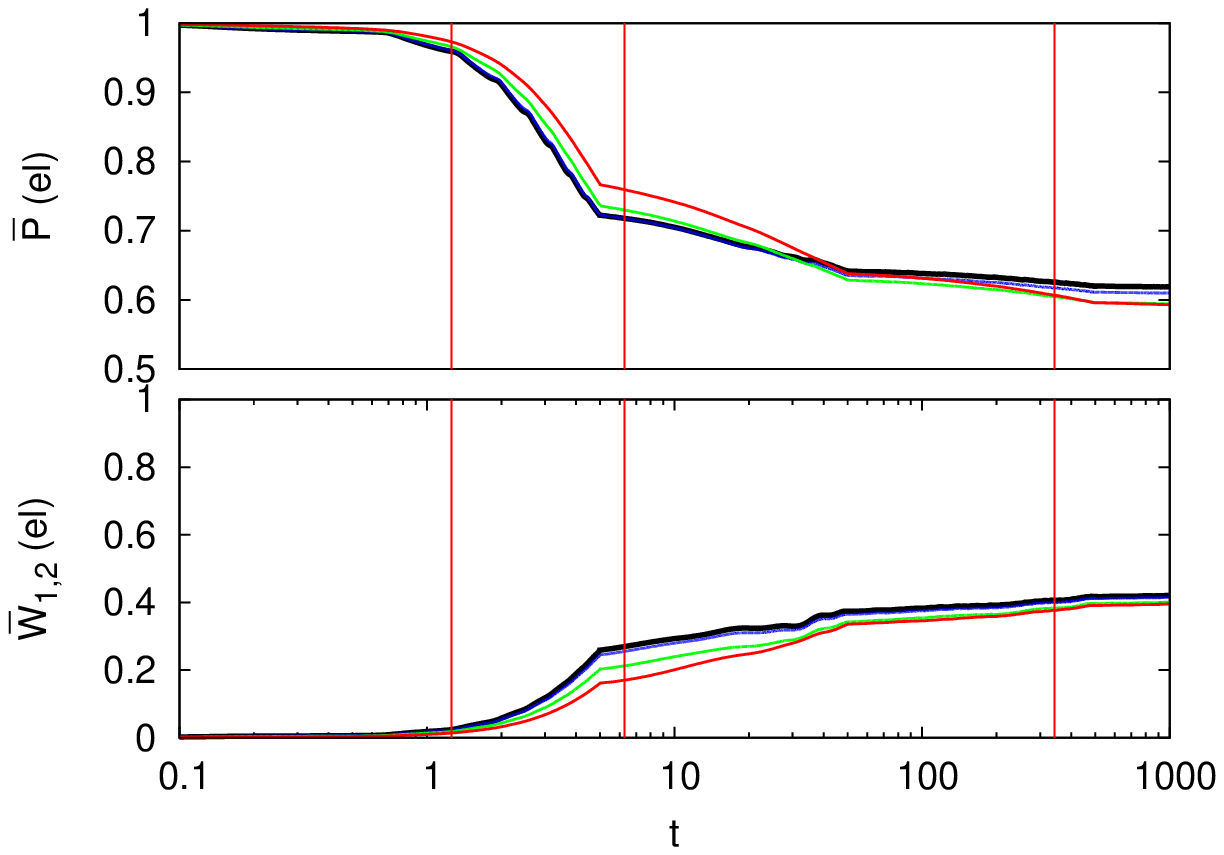}
\includegraphics[width=\figWc,angle=-0]{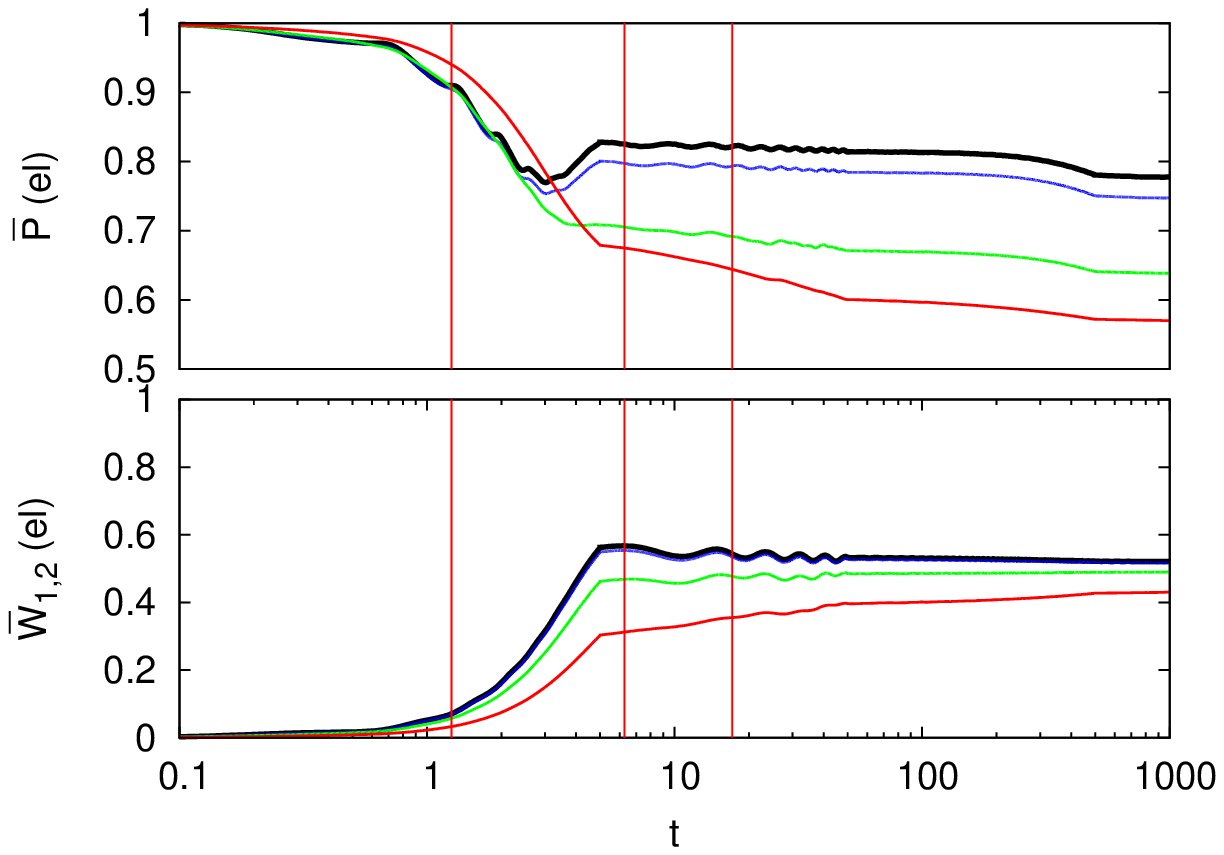}
\end{center}
\caption{Polaron (top) and electron (bottom) time-averaged populations and purity. 
Parameters and labels are the same of figure \ref{fig:anti}.
}
\label{fig:avanti}
\end{figure}

In figure \ref{fig:anti} is shown the time behaviour of the purity $P$ (Eq.
(\ref{purity})) as well as the transition probability ( (\ref{trans})) 
obtained in
the antiadiabatic regime when the phonon frequency ($\omega_0$) is much larger
than electron hopping $J$ for both (el) (\ref{eqn:elprep}) and (pol) 
(\ref{eqn:polprep}) 
initial preparations. Timescales defined in section
\label{sec:analytics} are shown as vertical lines, the time scale is logarithmic
to better show the much different time domains. We consider two parameters
sets at several temperatures. One characteristic of strong coupling (left
panels) and the other of weak coupling (right panel). The same sets of
parameters and temperatures is used in figure \ref{fig:avanti} where with show the
time-averaged $P$ and $w$. Let us first discuss the strong coupling regime.

It is known that, in the antiadiabatic regime,  the  polaron is a well
defined quasi-particle at  strong coupling \cite{lang}, in the sense that, in
the polaronic spectral function, almost all the spectral weight is contained in
the  polaronic peak.  This has  also been shown for a
two-site model  \cite{paganelciuk,ranninger,demello,demello1,robin,alexandrov}.
On the contrary, in the electron spectral function, the total spectral weight 
is distributed between a large number of frequencies \cite{paganelciuk}. 

From the point of view of transition probability and purity, the strong 
dependence on initial preparation can be seen comparing 
the low temperature evolution
for both the actual (figure \ref{fig:anti} left panel) and the
time-averaged (figure \ref{fig:avanti} left panel)
quantities. 

Let us consider the electron preparation (Figs. \ref{fig:anti} \ref{fig:avanti}
bottom left panels). Here the coupling with the oscillator mode is strong and the
system starting from a disentangled state (equation (\ref{eqn:elprep})) evolves in a
state in which electron is  entangled with the oscillator. This is shown  in the
purity evolution where we see that electron loses coherence very rapidly, on a
time scale $\tau_J$, and becomes a mixed ensemble, even at zero temperature.
\footnote{Notice that, in this case, the analysis of purity and transition
probability alone in principle do not allow to determine which states the mixture
is composed of (pointer states). In particular whether the states are localized
or not. However, a straightforward analysis of non diagonal elements of reduced
density matrix shows that the states are indeed localized.}

On the same timescale, transition probability approaches $1/2$ in average  (figure
\ref{fig:avanti} bottom left panel). The initial decoherence  is almost
independent on the temperature, as can be seen from averaged quantities,  while
long time recoherence peaks are suppressed as $T$ increases. 
Such a suppression results from destructive interference between time evolution
of the different terms appearing in  (\ref{rhoel1}) when excited oscillator
states are {\it initially} populated. 
Referring to the spectral analysis \cite{paganelciuk} and to figure \ref{fig:spectral}, this
phenomenon must be ascribed to the  superposition of a large number of high
frequency excitations.

 Even if the transfer does not have a regular 
shape, one can see some high frequency oscillations of period $\tau_{\omega_0}$.
These frequencies correspond to the energy separation between two  adjacent
electronic bands \cite{paganelciuk}.  

On the contrary, the  polaron preparation 
(see figure \ref{fig:anti},  top left
panel) evolves in a state which is 
completely coherent at zero temperature. The frequency associated
to polaron transfer is equal to the renormalized tunnelling  $J^*$, as predicted by
the HLFA (see  \ref{eqn:HLFAapprox}).  So, the state is pure and
delocalized.  The  polaron state remains coherent even for temperatures
comparable  with $\omega_0$, but higher frequency modulation appears  making the
state oscillate  from a  pure to a mixed one. Nevertheless, it is possible to see
an overall modulation of the transition probability {\it with the same period $\tau_Q$} even at
the largest temperature. This is in contrast with the HLFA at $T\neq 0$
(\ref{eqn:HLFAapproxT}) which  predicts that the  polaron band 
decreases with temperature and consequently $\tau_Q$ increases. However the purity
decreases as temperature increases,  as shown in figure \ref{fig:anti}. This is an
effect of the {\it broadening} of the polaron band that is observed as
temperature increases \cite{paganelciuk}. Indeed, a distribution of spectral
weight  among several poles around the polaron band occurs as an  effect of
increasing vibronic  excitations (ref. \cite{paganelciuk} figure 3 upper panel).
This leads to a  decoherence effect due to destructive interference between
these oscillating contributions to purity (see Eqs.
(\ref{eqn:risolventiel},\ref{eqn:rhopolFG})).  For high temperature
($T\gg\omega_0$), the state becomes completely mixed and the evolution of the
polaron is  analogous to that of the electron. This is  evident from
the highest temperatures curves shown in figure \ref{fig:avanti}, left panels upper and bottom
left. We conclude that the main source of
decoherence is temperature for polaron, while the electron decoheres even at zero temperature
due to the coupling with the vibronic mode.

This is also found in the weak coupling regime (electron preparation Figs.
\ref{fig:anti},\ref{fig:avanti} bottom right panels). Here electron coherence
approaches a value which is larger in average than that at strong coupling (figure
\ref{fig:avanti}) and decreases as temperature increases. 
We see (Figs. \ref{fig:anti},\ref{fig:avanti}  right
panels) that  the  polaron purity differs qualitatively from the
electron only near zero temperature, while averaged polaron and electron
transition probabilities is essentially the same for all showed temperatures. 
Indeed,  a sufficiently weak interaction is not able, at zero temperature, to
excite many  vibrational states,  so the electron decoherence is  essentially
given by  the small perturbation of the lowest oscillator's states.  On the
other hand, the  polaron is not formed (we are below the polaron
crossover) and the charge does not acquire much coherence by moving with the
oscillation cloud. Spectral analysis shows (ref. \cite{paganelciuk} figure 3
lower panel) that in weak coupling HLFA is qualitatively recovered, we have a
polaron band {\it narrowing} as temperature is raised up in contrast with the
strong coupling behaviour. 

It is worth stressing that, in both weak and strong coupling regimes, at high
temperature, the increasing number of  oscillator states involved in the initial
state produce decoherence on timescale 
$\tau_{J}$. 
Decoherence can be partial but nonetheless no environment is needed to explain
the decoherence process. The only source of decoherence are the states
populated by the initial thermal distribution 

\subsection{Adiabatic regime}

\begin{figure}[htbp]
\begin{center}
\includegraphics[width=\figWc,angle=-0]{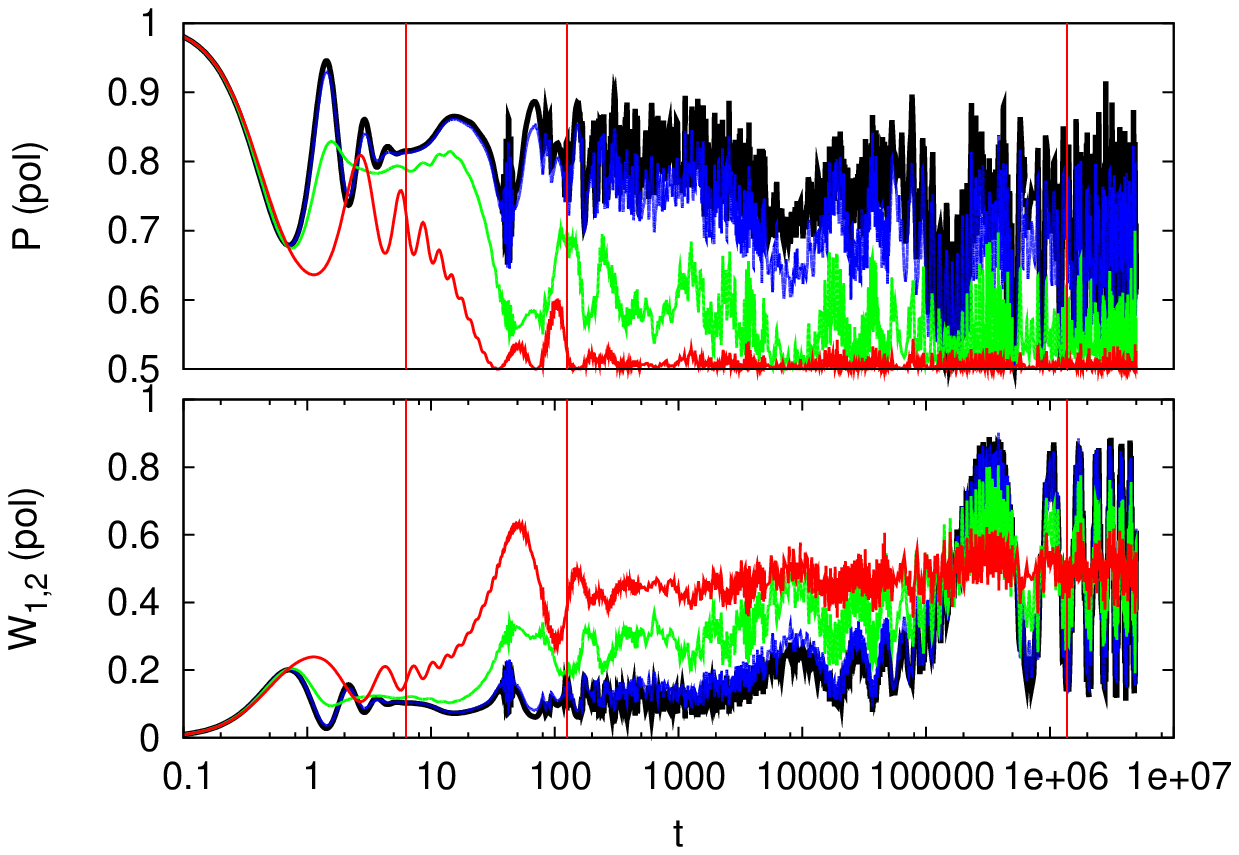}
\includegraphics[width=\figWc,angle=-0]{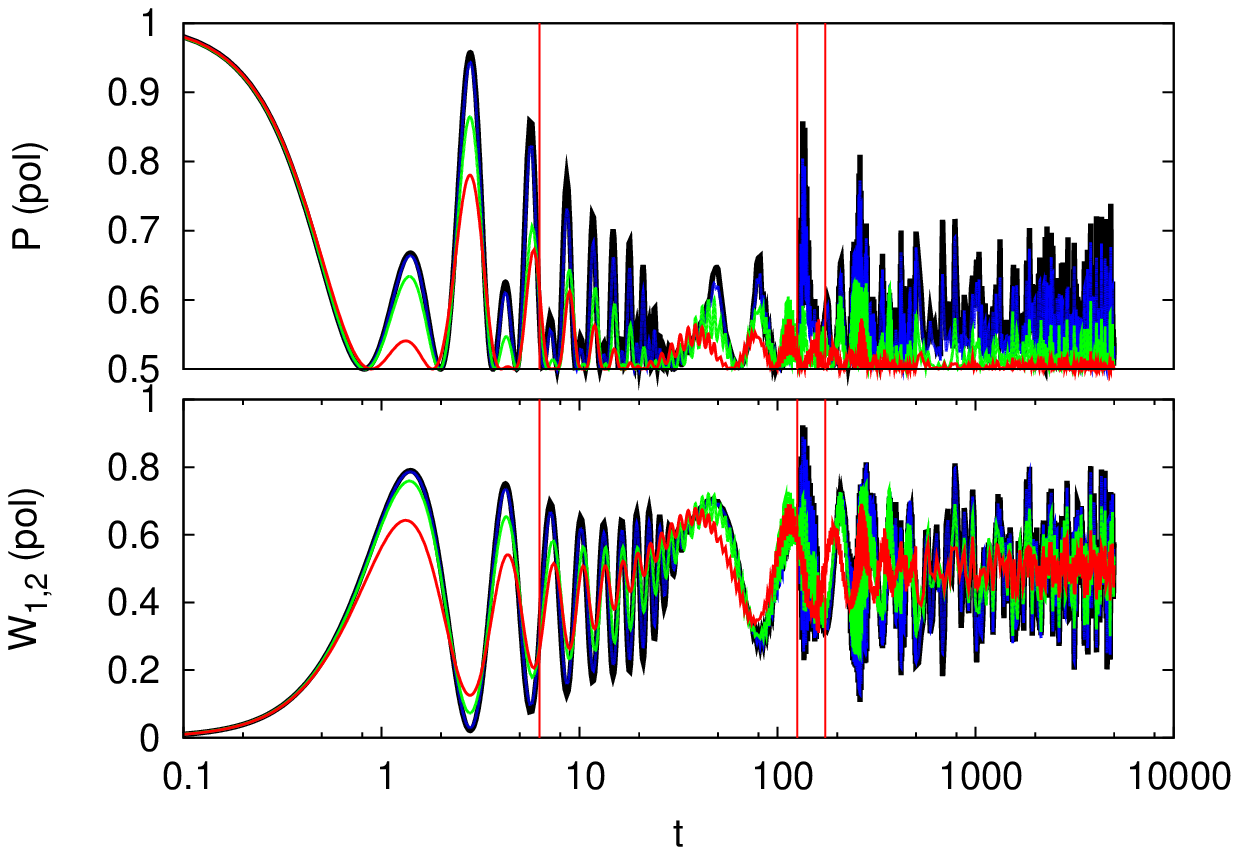}
\end{center}
\begin{center}
\includegraphics[width=\figWc,angle=-0]{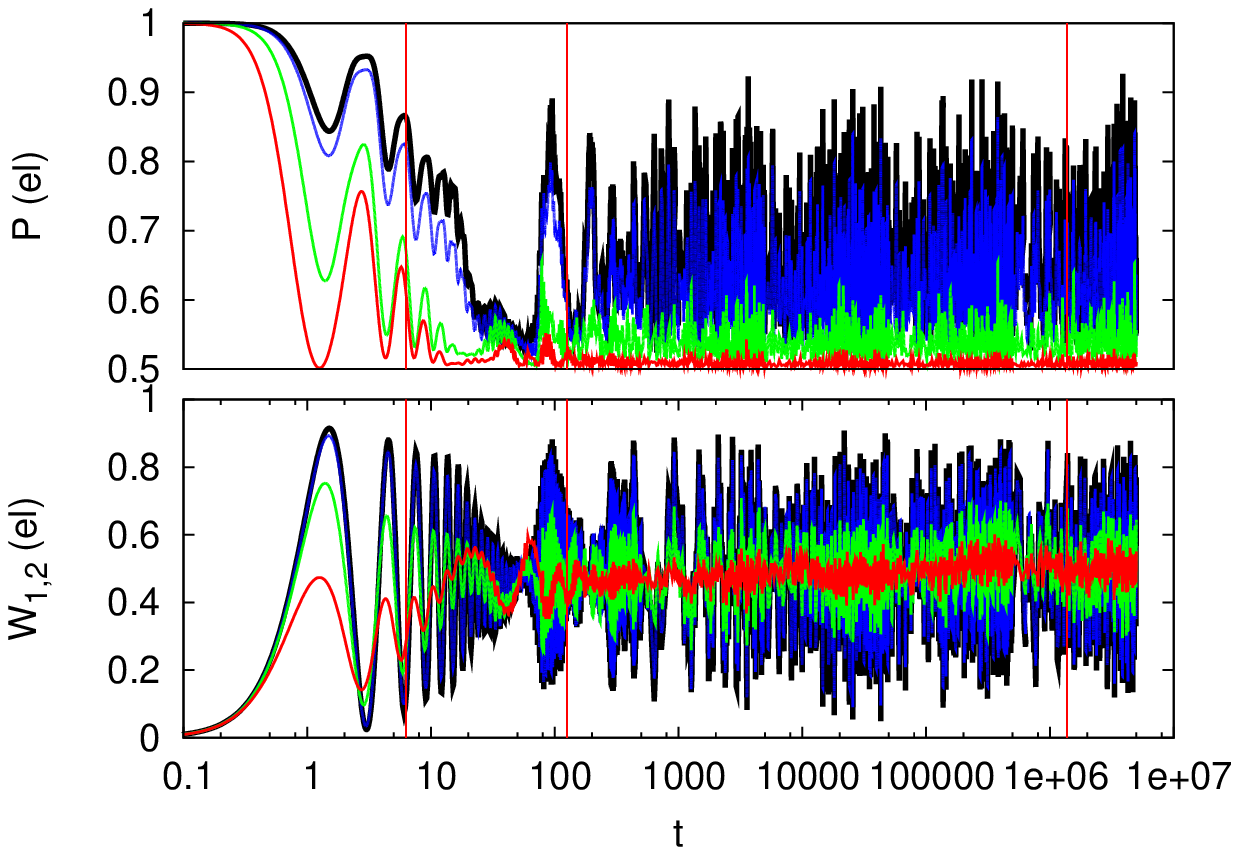}
\includegraphics[width=\figWc,angle=-0]{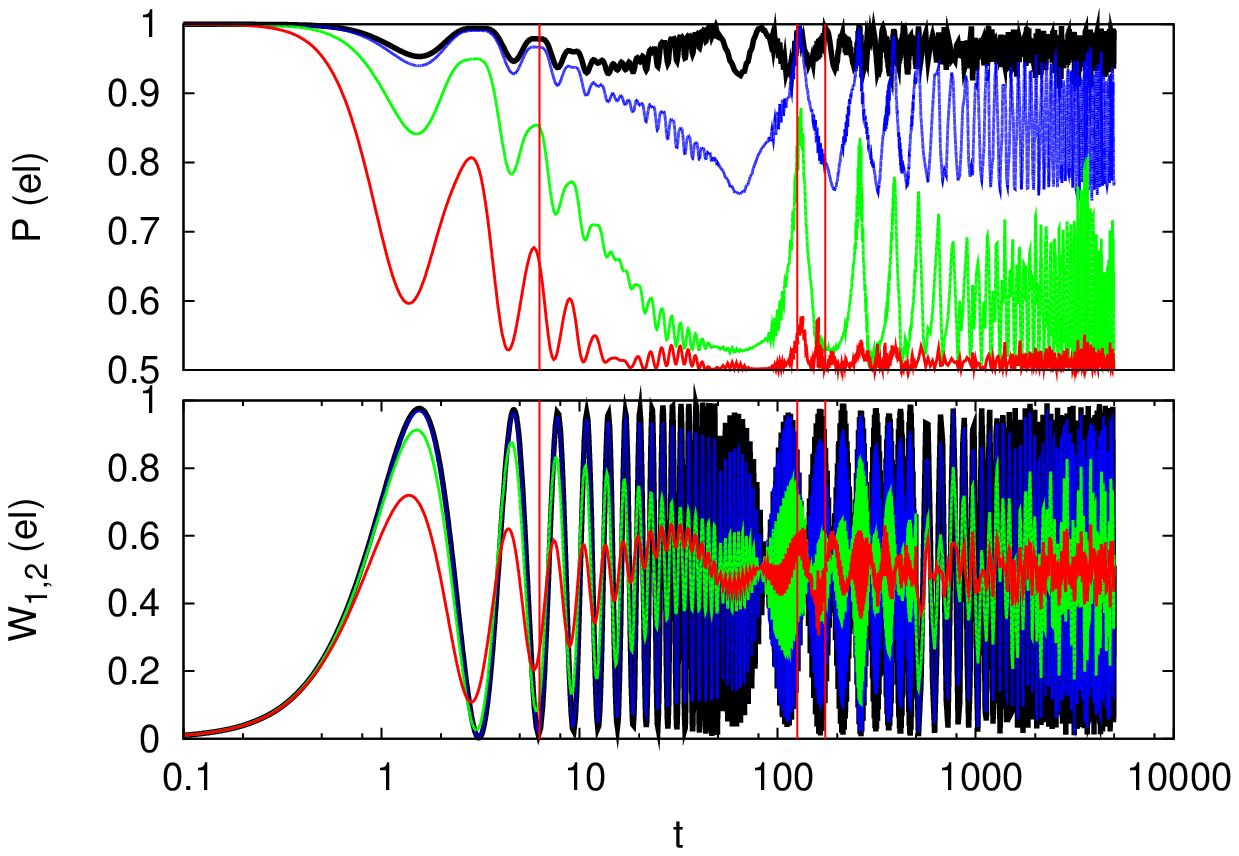}
\end{center}
\caption{Polaron (top) and electron (bottom) populations and purity. Left panels 
Adiabatic and strong coupling regime: $\gamma=0.1$ $\lambda=2$.
Right panels adiabatic and weak coupling regime $\gamma=0.1$ $\lambda=0.5$.
Curves are for $T/\omega_0=0.0$ (black), $T/\omega_0=0.5$ (blue),
$T/\omega_0=2.0$ (green), $T/\omega_0=10.0$ (red).Vertical lines marks from
left to right the timescales $\tau_J$,$\tau_{\omega_0}$,$\tau_Q$.}
\label{fig:adi}
\end{figure}

\begin{figure}[htbp]
\begin{center}
\includegraphics[width=\figWc,angle=-0]{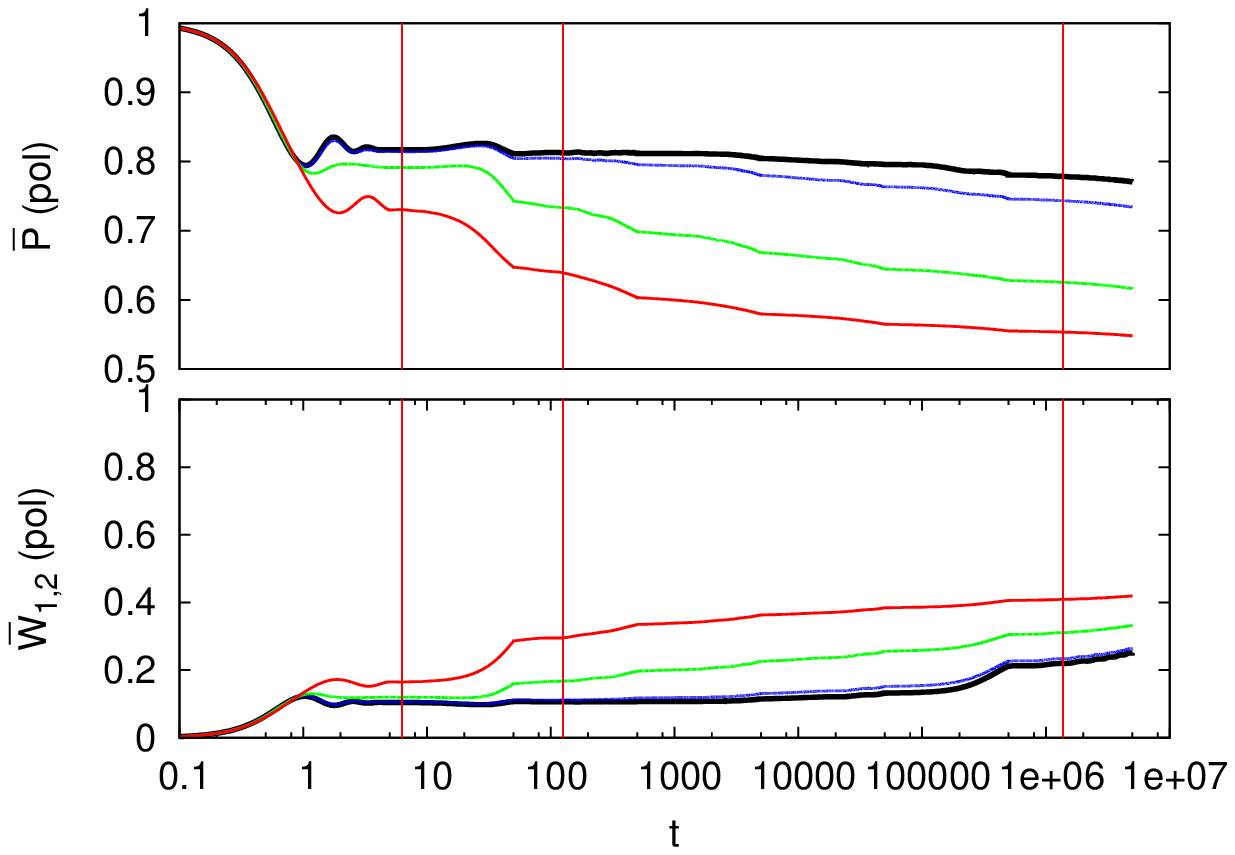}
\includegraphics[width=\figWc,angle=-0]{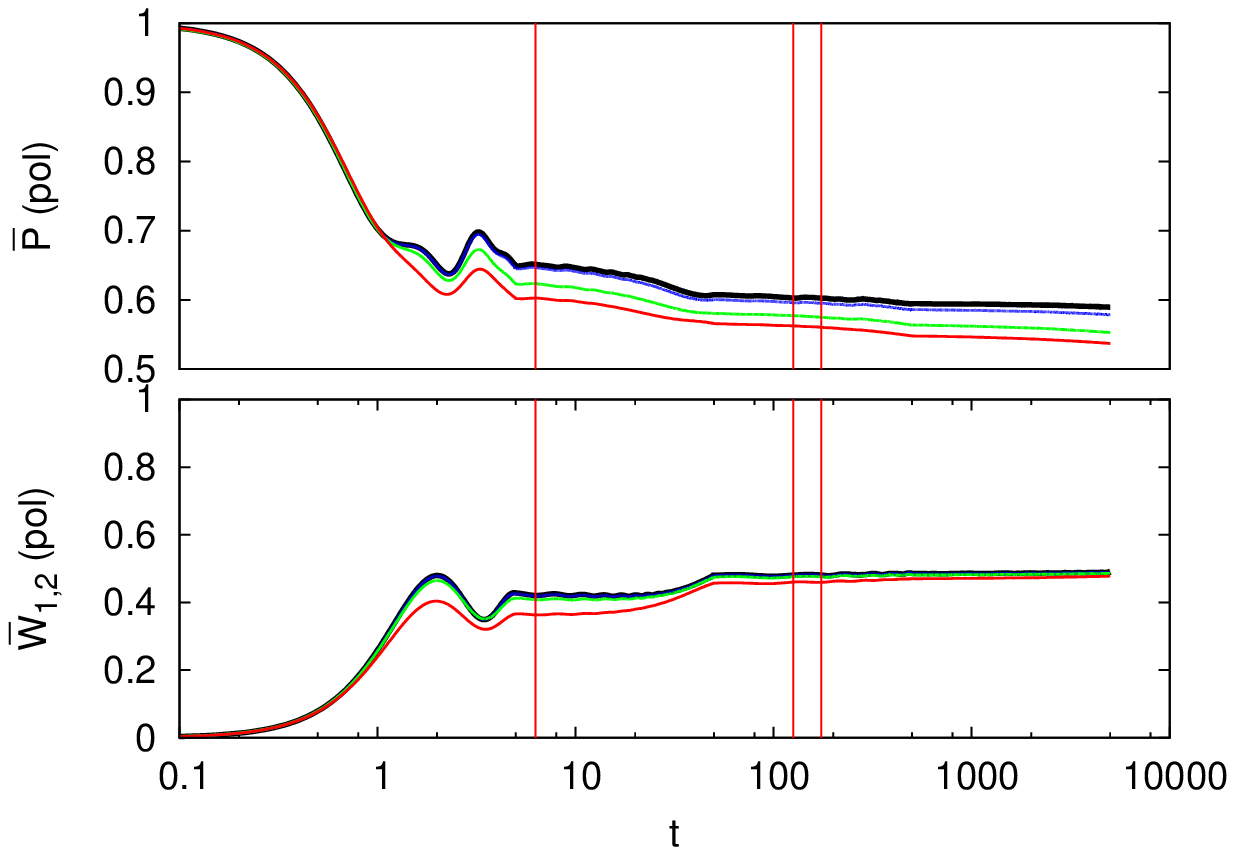}
\end{center}
\begin{center}
\includegraphics[width=\figWc,angle=-0]{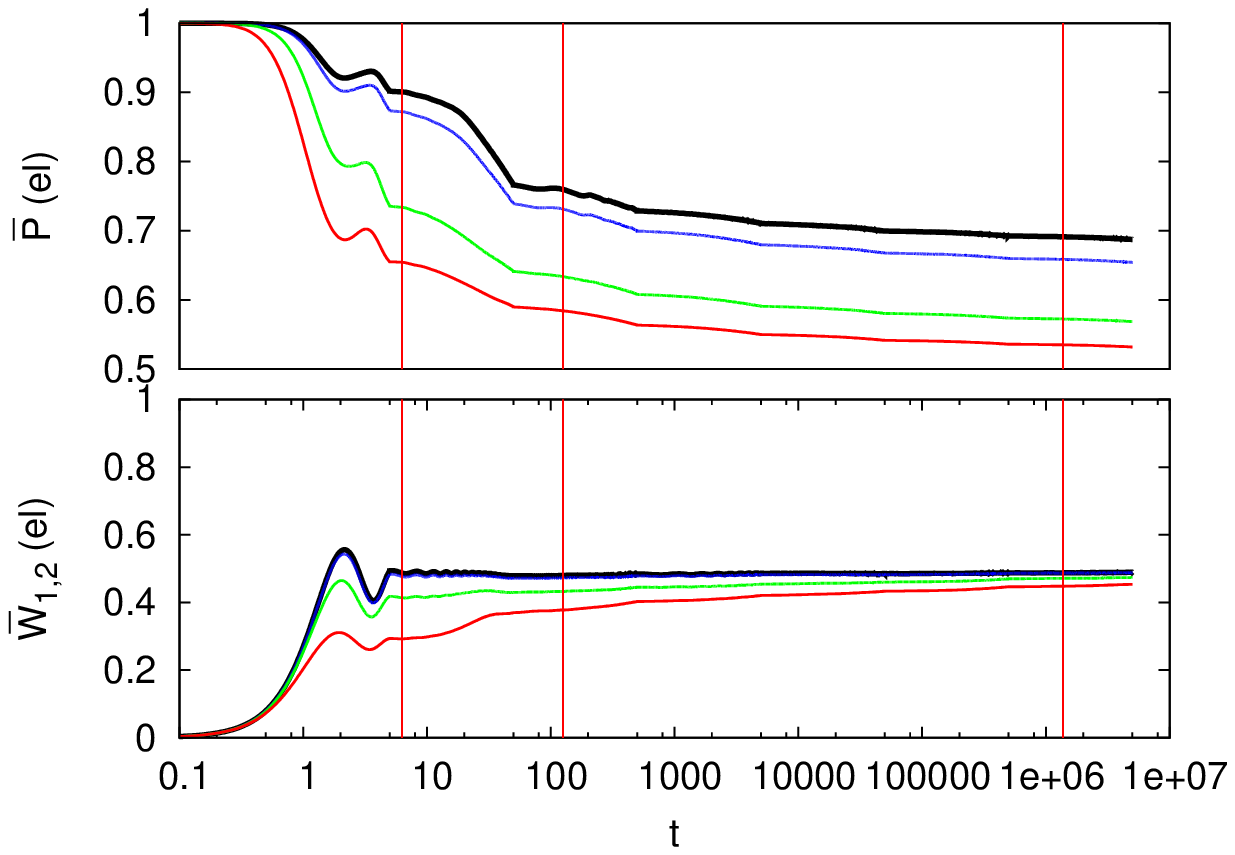}
\includegraphics[width=\figWc,angle=-0]{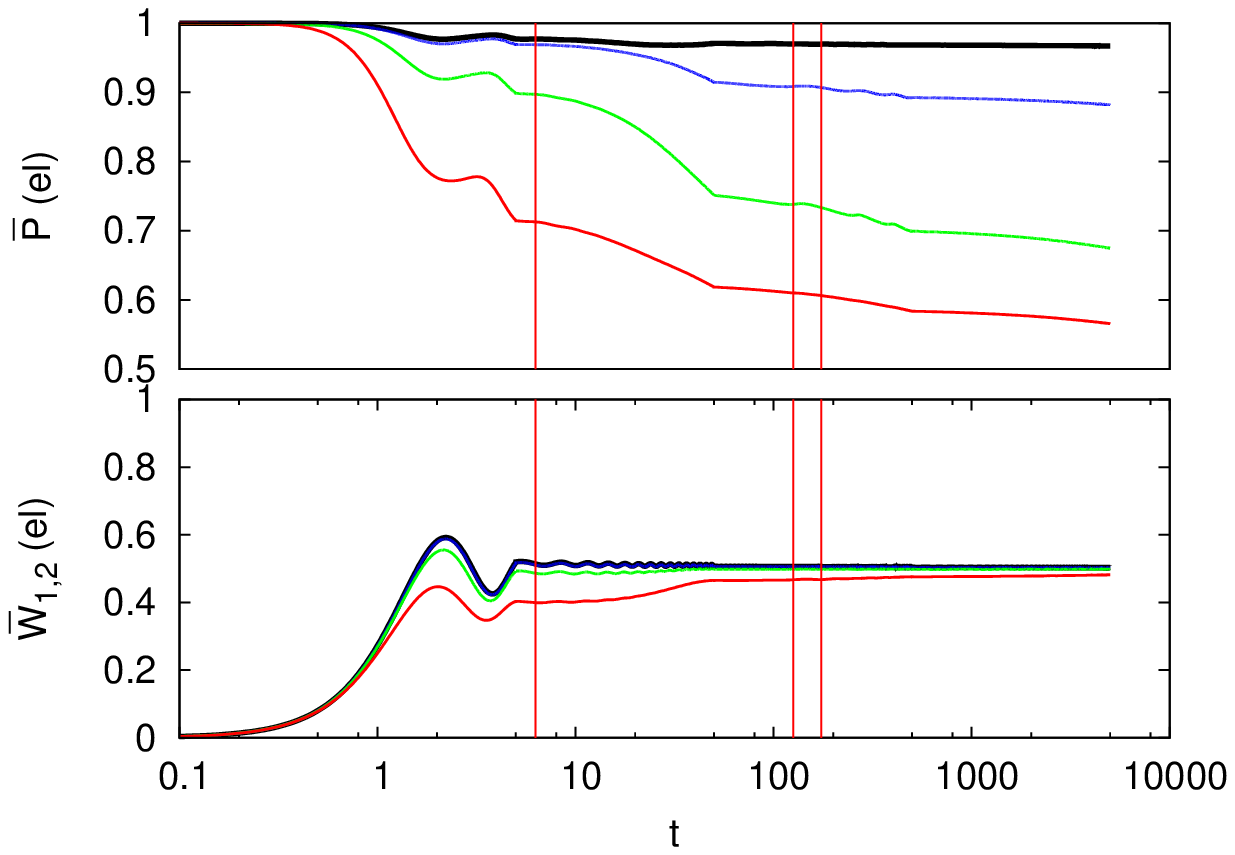}
\end{center}
\caption{Polaron (top) and electron (bottom) time-averaged populations and purity. 
Parameters and labels are the same of figure \ref{fig:adi}}
\label{fig:avadi}
\end{figure}

Results form the Exact Diagonalization method is reported in figure
\ref{fig:adi}, the averaged quantities are shown in 
\ref{fig:avadi} in the same way we did in the antiadiabatic case.
Notice that now the shortest timescale is $\tau_J$.

Let us first discuss the strong coupling case.
We see that, in contrast with antiadiabatic regime, there is a marked 
dependency on temperature of 
 {\it both}  electron and polaron properties. More specifically, 
polaron preparation no longer 
evolve coherently at low temperature. 

In the  first time scale, $\tau_J$, the particle is localized (its
transition probability is extremely low) but the state keeps on being quite pure. The
 polaron is trapped inside the initial site and 
both transition probability and coherence
evolve initially with characteristic time $\tau_J$ independently on temperature. 
At intermediate timescale $\tau_{\omega_0}$  temperature induces delocalization while the coherence
decreases.  In this time regime,  the  polaron transition probability is
related to the quasi classical motion of the oscillator and depends strongly on
temperature.

\begin{figure}[htbp]
\begin{center}
\includegraphics[width=\figW]{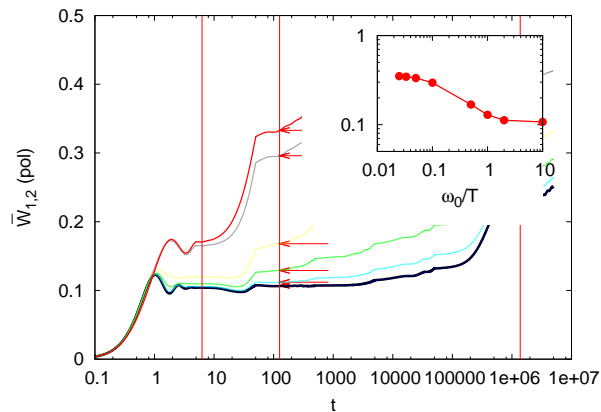}
\end{center}
\caption{Time averaged 
transition probability for $\lambda=2.0$,$\gamma=0.1$, temperatures are from
bottom to upper curves $T/{\omega_0}=0.1,0.5,1.0,2.0,10.0,20.0$. Vertical lines marks
from left to right timescales $\tau_J$,$\tau_{\omega_0}$ and $\tau_Q$ respectively.
Inset: levels reached at time $\tau_{\omega_0}$ (arrows in the main panel) 
as a function of the inverse
temperature.}
\label{Arrhenius}
\end{figure}

This can be seen in figure \ref{Arrhenius}, where we plot the temperature
dependence of the level reached by the averaged transition probability on the
time scale $\tau_{\omega_0}$ (inset). Since there is no clear plateau in the
averaged transition probability for times greater than $\tau_{\omega_0}$,  the
choice of the transition probability level it is rather arbitrary. We choose the
value of $\bar{w}_{12}$ at $\tau_{\omega_0}$. We see  that
this transition probability level  pass from a low temperature  behaviour, which
is temperature independent, to a temperature dependent behaviour trough a wide
crossover. 

At very low temperature, after the characteristic time $\tau_{\omega_0}$,
coherence and transition probability reach a quasi stationary value that is
essentially dominated by fast tunnelling of the charge between the two sites, 
with a given phonon displacement.
Once temperature increases, classical activation processes  of the phonon coordinate
becomes effective, producing an increase in transition probability as well as
a decrease of the purity. As we shall see in the next section, this thermal activated behaviour
is to be ascribed to the phonon classical hopping between two adiabatic minima and disappears
in the SA where such hopping events are absent. 

As far as  electron is concerned,	we can see that the site occupancy begins
to  oscillate coherently, with period $\tau_J$, with a damping increasing 
with temperature. In the same timescale, the time averaged 
value shows a saturation at low temperature.
At a {\it temperature independent} intermediate  timescale $\tau_{\omega_0}$,
time averaged  coherence reaches a very slowly  decreasing level which
decreases with increasing temperature. 
Quantum oscillations still exist
further in time, but the remaining  coherence slows down  at high temperature, so
that the long timescale  ($\tau_Q$) seems to be not relevant in this case.
The electron's tendency to coherently hop is suppressed by decoherence, induced
by excited phonons, whose number increases with temperature.
In the adiabatic limit, this effect is more evident than in the antiadiabatic
case, this is  because the energy spacing between the oscillator's levels
becomes very small and the spectrum tends to a continuum.

In the weak coupling case (figure \ref{fig:adi} and  \ref{fig:avadi} right panels) 
the transition probability is very similar for both
electron and polaron as in the antiadiabatic case. 
Conversely, at low temperature, the electron is much coherent than 
the polaron. In this regime, being the adiabatic potential single-well, the displaced 
phonon base is not the best choice. So, many displaced oscillator states are
coupled with polaron, which decoheres rapidly.

\subsection{Comparison with quanto-classical approaches}


\begin{figure}[htbp]
\begin{center}
\includegraphics[width=\figWc,angle=-0]{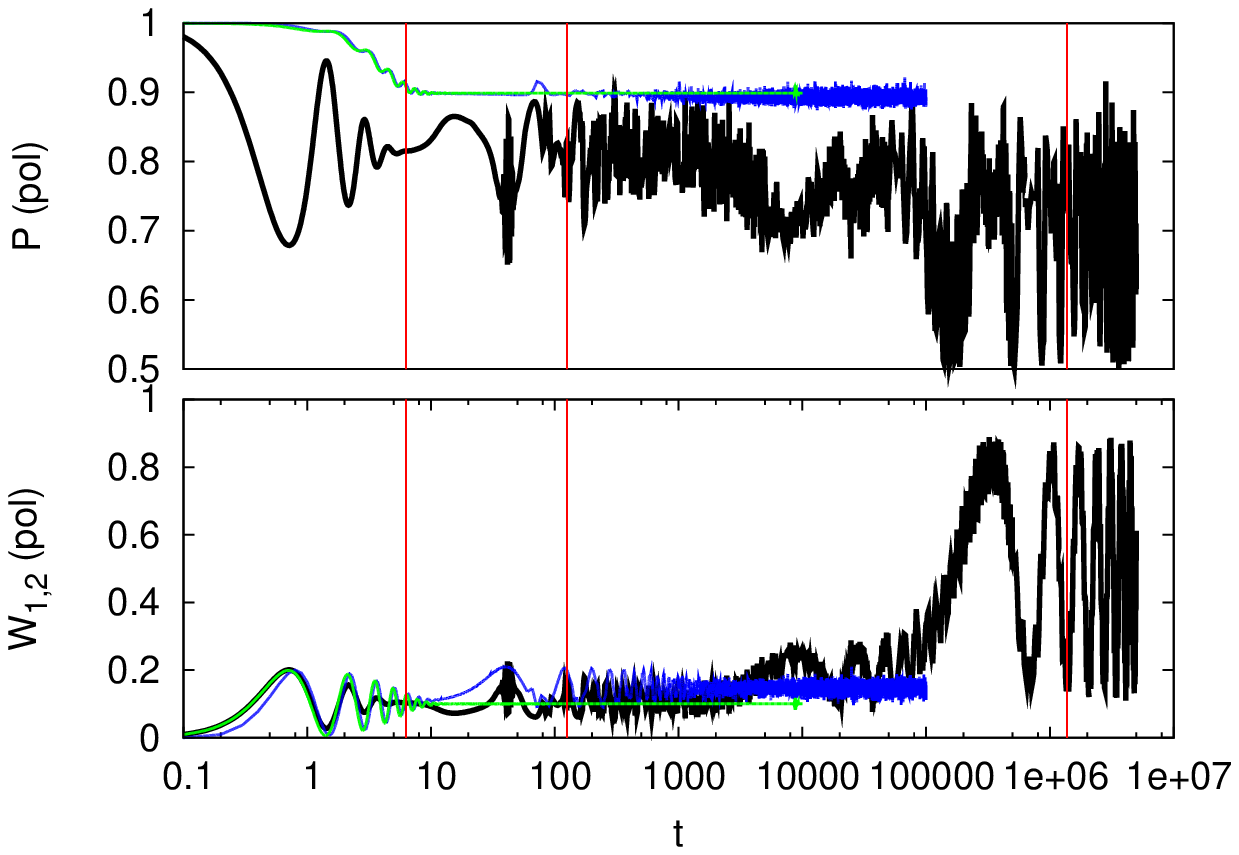}
\includegraphics[width=\figWc,angle=-0]{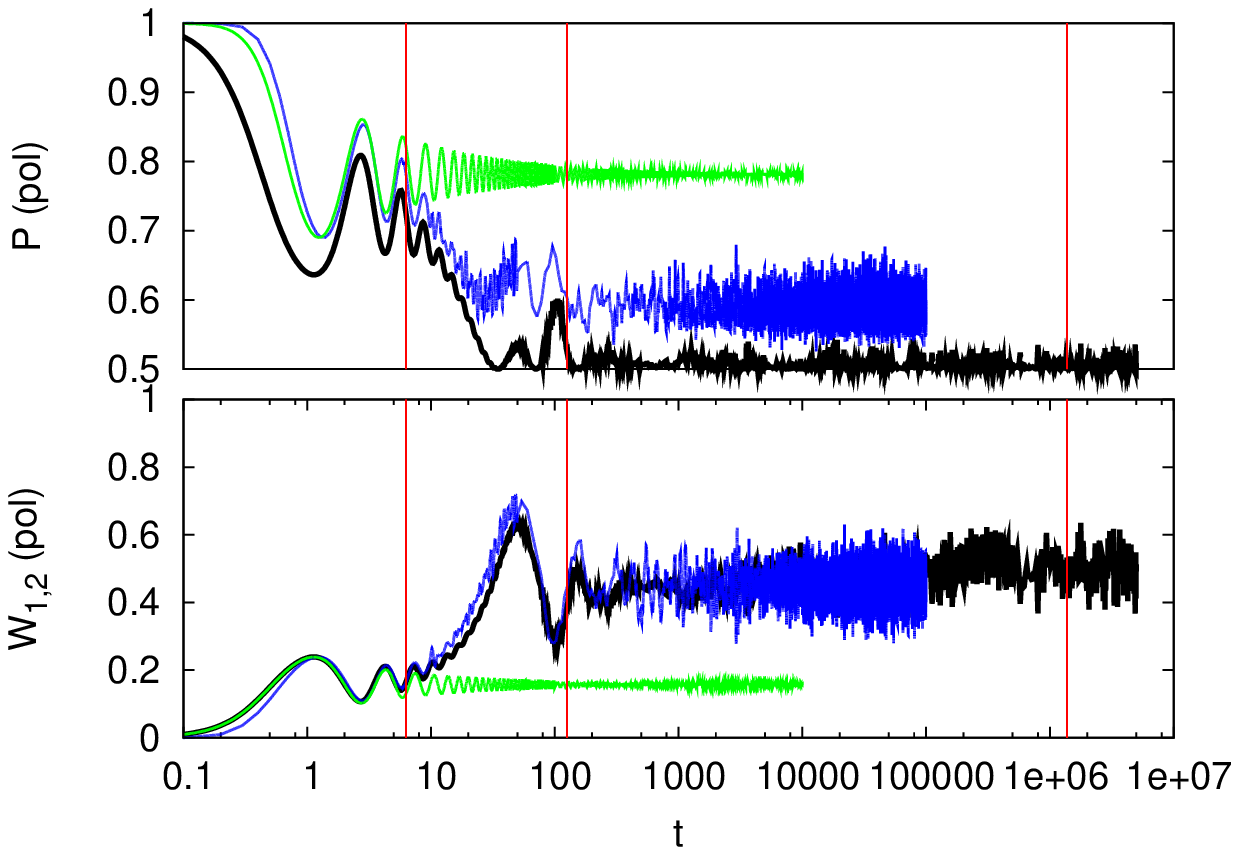}
\end{center}
\begin{center}
\includegraphics[width=\figWc,angle=-0]{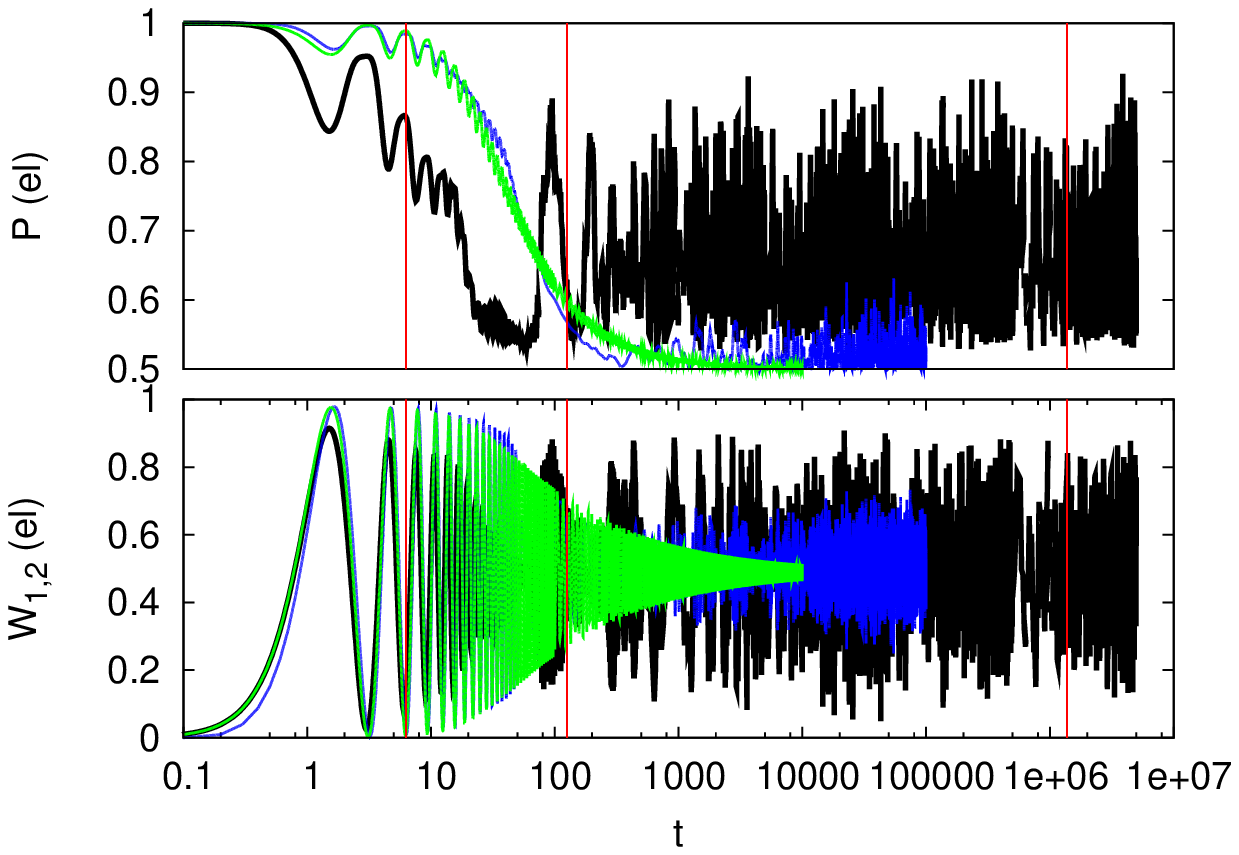}
\includegraphics[width=\figWc,angle=-0]{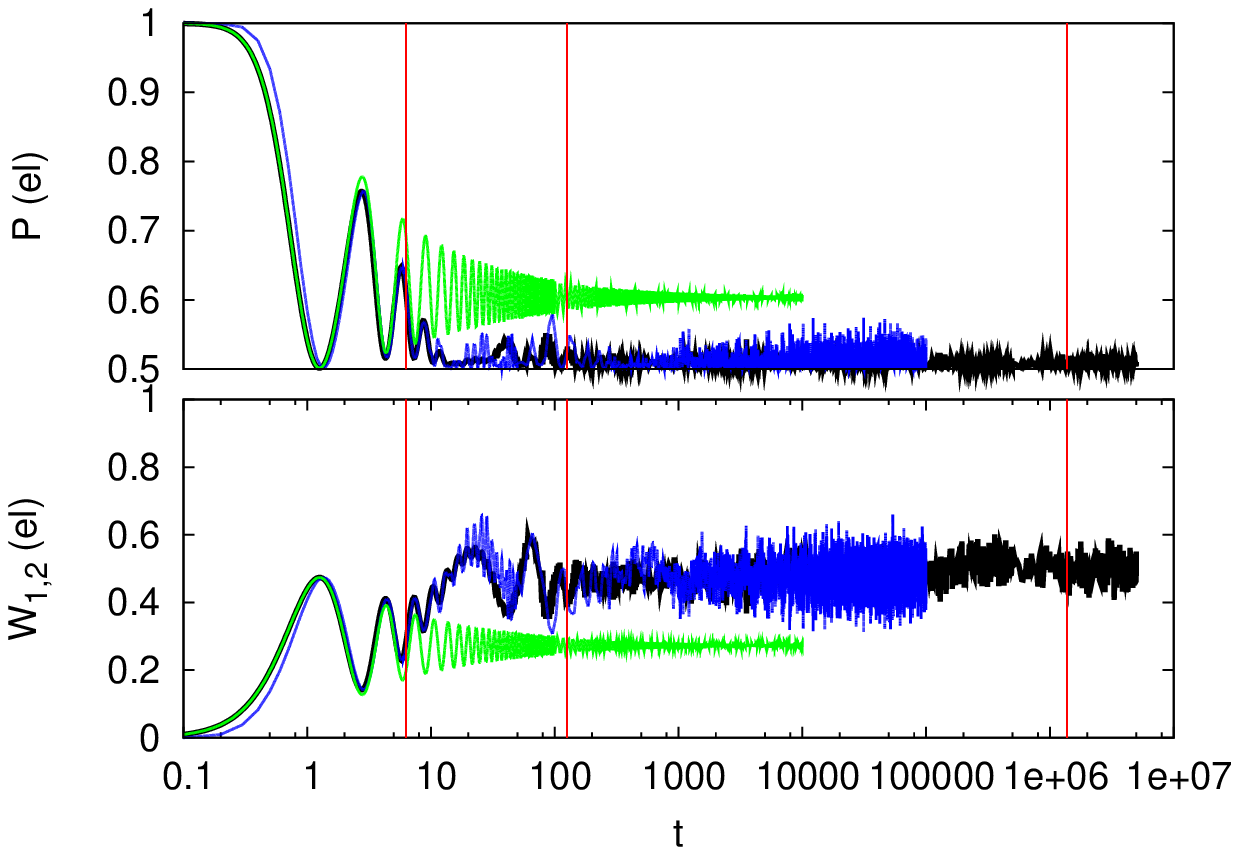}
\end{center}
\caption{Polaron (top) and electron (bottom) populations and purity in the adiabatic
strong coupling regime $\gamma=0.1$ $\lambda=2$. Left panels 
low temperature $T/\omega_0=0.1$.
Right panels high temperatures $T/\omega_0=10.0$.
Curves refers to ED (black), QC (blue),
SA (green) approximations.
Vertical lines marks from
left to right the timescales $\tau_J$,$\tau_{\omega_0}$,$\tau_Q$.}
\label{fig:cfr_adiab_strong}
\end{figure}

\begin{figure}[htbp]
\begin{center}
\includegraphics[width=\figWc,angle=-0]{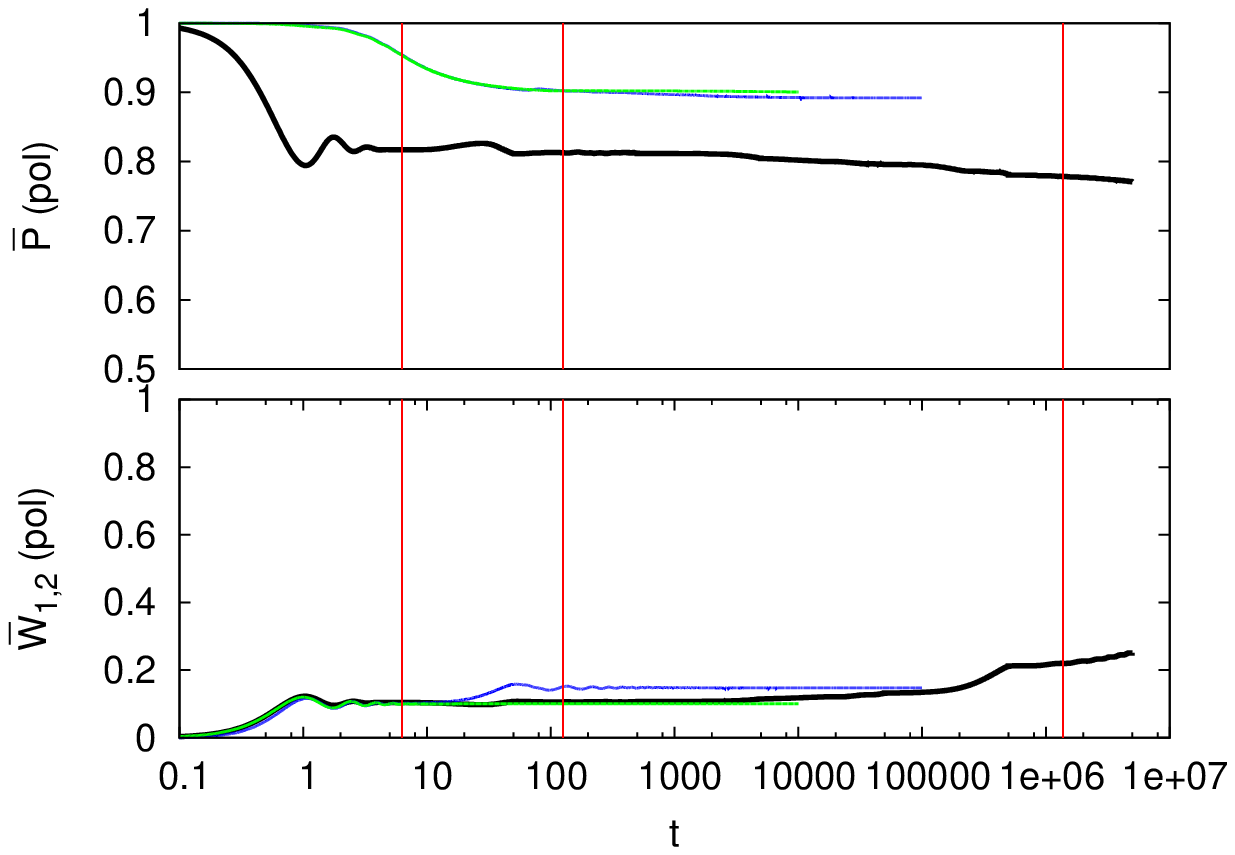}
\includegraphics[width=\figWc,angle=-0]{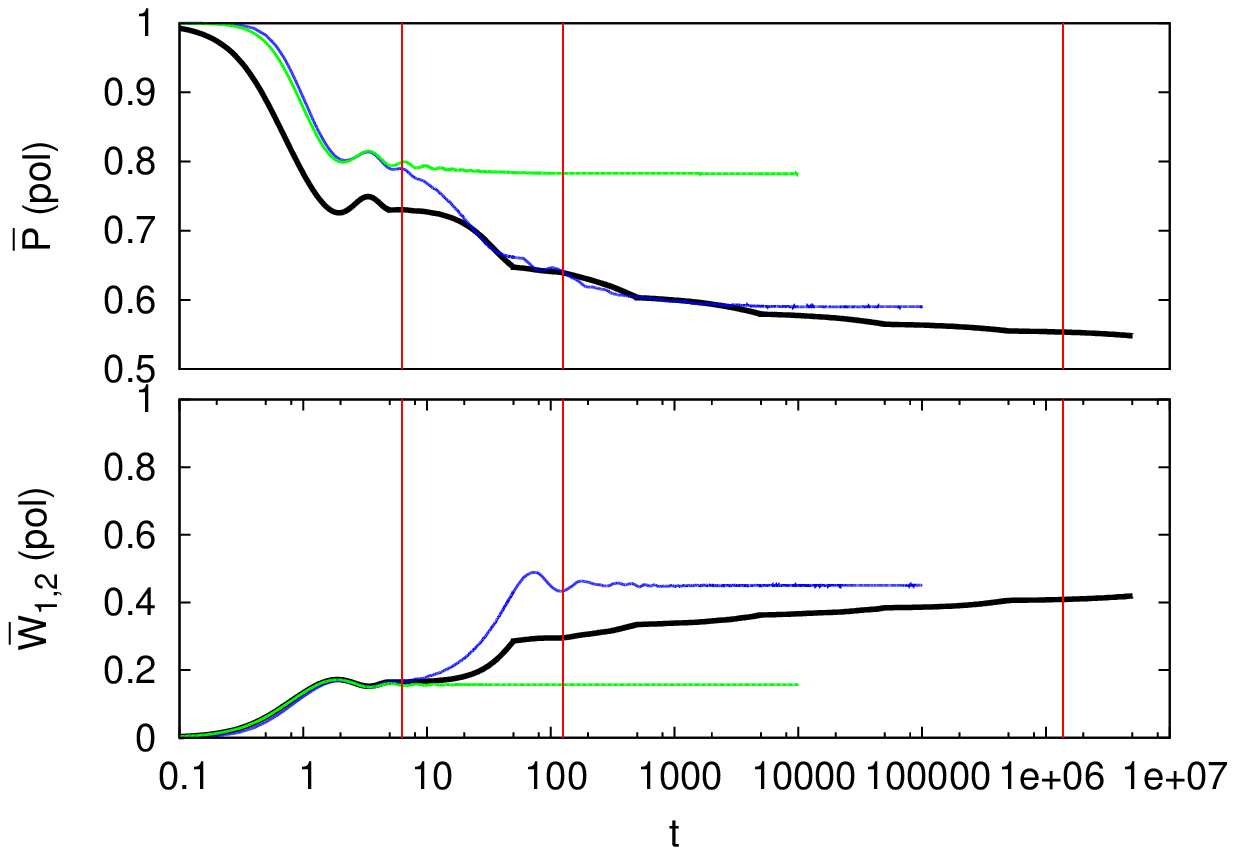}
\end{center}
\begin{center}
\includegraphics[width=\figWc,angle=-0]{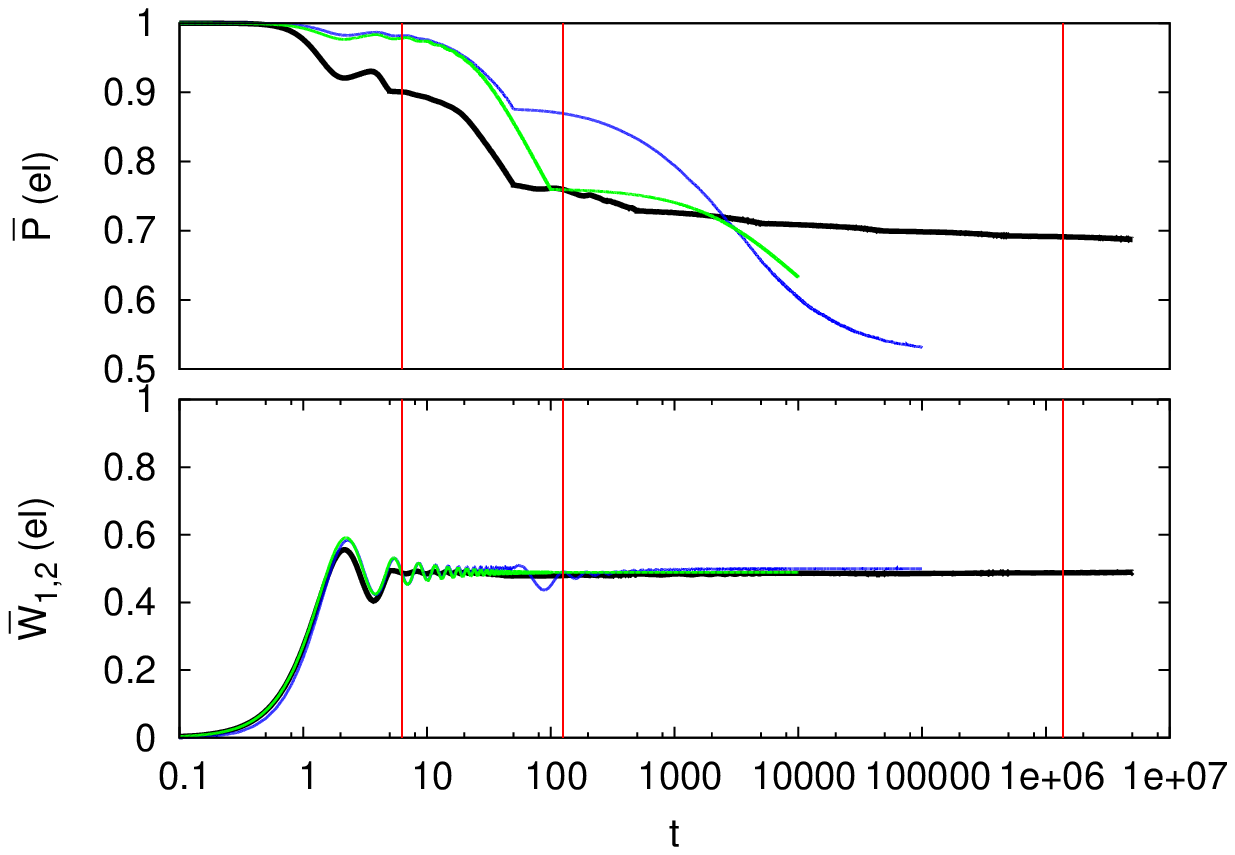}
\includegraphics[width=\figWc,angle=-0]{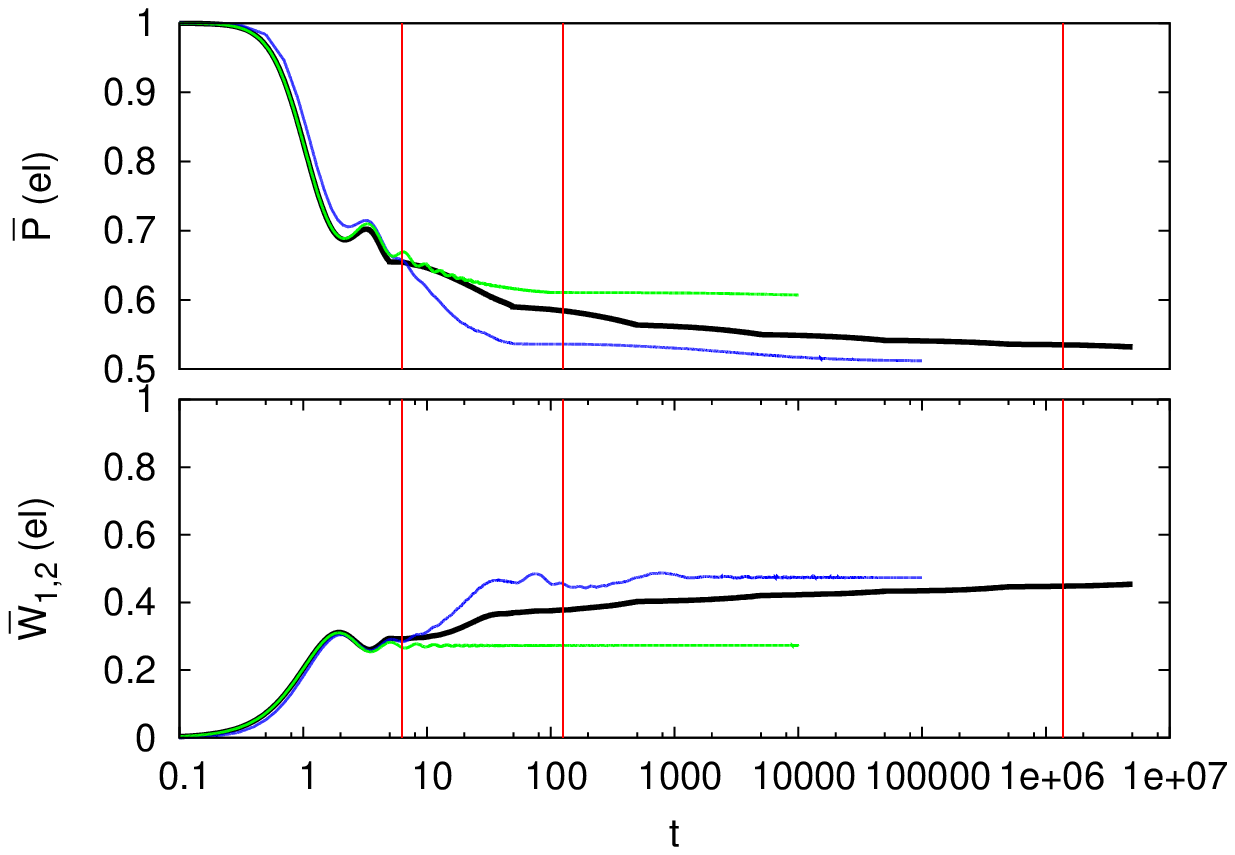}
\end{center}
\caption{Polaron (top) and electron (bottom) time-averaged populations and purity. 
Parameters and labels are the same of figure \ref{fig:cfr_adiab_strong}}
\label{fig:cfrQCTS}
\end{figure}

\begin{figure}[htbp]
\begin{center}
\includegraphics[width=\figWc,height=\figHc,angle=0]{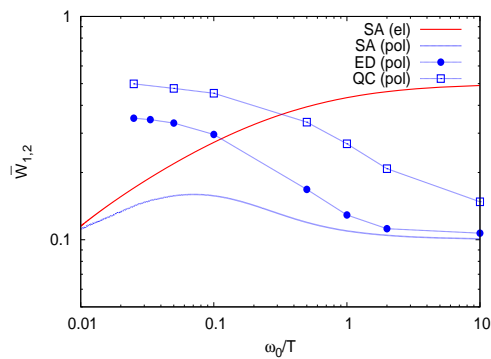}
\end{center}
\caption{Time average transition probability level at the timescale $\tau_{\omega_0}$
in the adiabatic regime 
($\lambda=2.0,\gamma=0.1$)
as a function of inverse temperature for different preparations and different approximations.}
\label{fig:av}
\end{figure}
In this section, we show a comparison between the results obtained in the 
three different ways described before: exact diagonalization (ED) by means of 
mapping introduced in Sec. \ref{sec:analytics}, the quantum-classical (QC) dynamics approach
described in Sect. \ref{app:QC} and the static (SA) approximation 
(Sect. \ref{sub:adiab}).
We shall limit ourselves to an adiabatic case ($\gamma=0.1$) 
with electron-phonon interaction strong enough to allow the polaron formation
($\lambda= 2$).


In figure \ref{fig:cfr_adiab_strong}, is reported
the exact dynamics given by the three different techniques, while 
in figure \ref{fig:cfrQCTS} is shown the time average. Remember that, 
as far as the polaron  is concerned, in both SA and QC approximation, the purity 
represents that of the
electron with a initially displaced phonon distribution.

At low temperature (left panels), and within the $\tau_J$ timescale, the classical 
phonon is almost freezed, and so both SA and QC approximation are equivalent. 
Nevertheless, the ED
behaviour, of both polaron and electron preparation, is quite different because of
 quantum fluctuations. In particular, for short 
timescales, one can see that the ED dynamics, at low temperature, is damped faster 
than the other two approximations. 
The difference becomes 
more evident for higher timescales.
In this regime, the temperature is not so 
effective in dissipation  processes, while the strong coupling and the quantum 
uncertainty produce a sort of purely quantum damping. 

A damping of this sort was also found in ref. \cite{lucke},
where authors consider a spin-boson Hamiltonian for a tunneling system coupled to a 
multi-mode bath with a ohmic spectral density.
It is worth stressing that, in the present case, 
such a damping is not due to the bath but rather a consequence
of the entanglement between the tunnelling  system and the quantum oscillator.
Nevertheless our results are comparable with that of ref. \cite{lucke}.
The reason is that, at strong coupling, the charge is coupled to 
the bath through a single collective mode, taking into account 
all the bath's oscillators.

As expected, the exact polaron purity completely differs from the purity
obtained by SA and QC. The reason is that, at low temperature and strong coupling, 
the polaron is well defined and moves as a quite coherent particle, as can be seen from the long timescale oscillations.
On the contrary, if we consider only an electron with a displaced oscillator, 
the particle remains localized because  the trapping 
mechanism, but it cannot coherently tunnel between the two sites.

A  semiclassical behaviour is approached for $T$  greater than 
$\omega_0$, when SA and  QC reproduce the ED transition probability
 within the $\tau_J$ timescale, as results evident in the 
right panels of Figs. (\ref{fig:cfr_adiab_strong}-\ref{fig:cfrQCTS}).
The QC approach remains a good description also for higher timescales.
Notice that the same occurs in the transport of extended system where classical incoherent transport 
is achieved when $T$ is greater than $0.2 \omega_0$ \cite{frat2}.
It is worth noticing that for very high temperature ($T \simeq J$,$T/\omega_0=10$ in figs. \ref{fig:cfrQCTS},\ref{fig:av}) the polaron is not formed,
its dynamics approaches that of 
the electron in an initially delocalized phonon distribution.
As a result, the QC purity approaches the ED's.

At high temperature, the oscillator dynamics plays a relevant role, QC is a much better approximation of ED than SA.
This fact can be understood by realizing that the main temperature effect is
the damping of the coherent tunnelling oscillations.
Once these oscillation are sufficiently suppressed, the phonon driven dynamics prevails.
In the SA framework, the initial thermal distribution of the phonon coordinate 
makes the electron thermalizes irreversibly in a time that is the shorter  the greater  the temperature.
Before this adiabatic thermalization, i.e, in a tunnelling period, the SA is still a good  approximation. 

Afterwards, the hopping of the oscillator coordinate into the other minimum
 of the  adiabatic potential (equation (\ref{eqn:adiabatic_energies})) takes place.
The charge degree of freedom follows while $w_{1,2}$ saturates on average. In extended systems this regime 
corresponds
activated mobility regime  \cite{kudinov,kramers,hanggi1990}. Since SA completely neglects the oscillator's dynamics,  it does not predict 
correctly  $w_{1,2}$, as can be seen in figure \ref{fig:av}.
The QC approximation, instead, gives a correct qualitative prediction. 

\section{Conclusions}
\label{sec:concl}

In this paper, we have studied a simplified model to treat the dynamics of a 
tunnelling charge interacting with  a vibrational degree of freedom. We
introduced a reduced density matrix approach to  characterize the  charge
dynamics. Temperature is introduced by taking an initial equilibrium
distribution of the oscillator. Both the transition probability and the purity
are studied, in order to connect the charge transfer with its coherence. 

Due to the simplicity of our model, we were able to span all the  parameter's
space even at high temperature and strong coupling and to study the role of
initial preparation. Moreover,  we can explore  a temporal range which is very
large, compared with the typical time scales that can be obtained in models
where the charge is coupled with a many degree of freedom oscillators' bath 
\cite{lucke,bonella2005,egger1994}.

As in any finite system, in our model, transition probability and purity can be
expressed as a superposition of many non commensurate by oscillations. We
therefore expect an oscillatory behaviour in our quantities of interests.
However the initial thermal distribution of the oscillator states  induces
decoherence on  intermediate timescale, due to the strong interaction with the
oscillator. This phenomena occurs depending on initial preparation of the
system.

We find that, in the antiadiabatic and strong coupling regime, the polaron 
exhibits a coherent tunnelling dynamics over time scales of the inverse polaron
renormalized  band. The coherent behaviour is lost out of the polaronic phase,
i.e. increasing the temperature or decreasing the coupling. Electron evolves
though partially incoherent dynamics. In the adiabatic strong coupling regime,
temperature enhances the incoherent polaron charge transfer. The opposite occurs
in the electron preparation. 

In the adiabatic regime, two common approximations has been compared with exact 
results, the aim is to highlight the limits of validity of these approximations
and to provide a simple testing tool, the two-site model, for generalizations to
other extended models. As expected, a dynamical semi-classical approximation
gives good estimates for both coherence and tunnelling amplitude, at high
temperature $T\gg\omega_0$. 
Quite unexpectedly, it allows for a good approximation of the transition
probability at low temperature, as far as time averaged quantities are concerned.
However, such a quasiclassical approximation fails approaching the anti-adiabatic
regime where non adiabatic transition are expected to contribute significantly
to charge dynamics. This simplified model could serve to test approximate
schemes to deal with this regime \cite{bonella2005}.

To conclude, we have shown that a {\it non dissipative} evolution of a tunneling
system, strongly coupled to a single oscillator, can give rise to decoherence
phenomena when the initial distribution of the oscillator is thermal and when
the oscillator distribution is {\it not} initially equilibrated in the presence
of the charge, that is  sufficiently far from the thermal equilibrium
distribution in the presence of the charge.  These decoherence phenomena are
independent on the presence of a dissipative bath. Thus, in a non equilibrium experiment in
which a charge is introduced in a molecular system and interacts strongly with a
particular mode of the molecular system, decoherence effects can be triggered
alone by this coupling and by the initial non equilibrium distribution of the
molecule.

\ack
Authors wish to thank Sara Bonella, Carlo Pierleoni and Sergio Caprara 
for fruitful discussions and suggestions.

\appendix
\section*{Appendix}
\section*{Atomic limit}\label{app:atomic}
In the atomic limit, the Hamiltonian is diagonalized by the
so-called Lang-Firsov (LF) transformation 
\begin{equation}\label{eqn:lfsep}
	D=\rme^{\alpha \sigma_z (a^\dag-a)}.
\end{equation}
This transformation shifts the phonon operators by a quantity $\alpha$, while
the electron operators are transformed in new fermionic operators, with energy $E_p$, associated to
a quasi-particle called  polaron\cite{tiablikov,holstein}.  This particle can be tough as a 
charge moving together with a dressing cloud of oscillator quanta,  $\alpha^2$ represents
the mean number of phonons in the  polaron cloud.

The atomic Hamiltonian 
\begin{equation}
H_0 = \omega_0 a^\dag a -\tilde{g}\sigma_z (a^\dag+a),
\end{equation}
after  the LF transformation
$\bar{H}_{0}=D^\dag H_0 D$ becomes:
\begin{equation}\label{eqn:AtomicLimit}
    \bar{H}_{0}=\omega_0 a^\dag
    a+E_p/2,
\end{equation}
the eigenvalues $E_n=\omega_0 n+E_p/2$ correspond to the two-fold degenerate
eigenvectors $ \ket{\psi_n^j,j}= D \ket{n,j}=\bar{c}_j^\dag \ket{n}$, were the
index $n=0,\ldots,\infty$ refers to the photon number, $j=1,2$ to the electron
site and $\bar{c}_j^\dag$ is the  polaron creation operator $\bar{c}_j^\dag=D
c_j^\dag D^\dag=c_j^\dag \exp\{(-1)^j \alpha(a^\dag-a)\}$.

In the case of finite $J$, the hopping term is not diagonalized by
(\ref{eqn:lfsep}) and  the new Hamiltonian $\bar{H}=D^\dag H D$
becomes
\begin{eqnarray}\label{eqn:parappa}
    \bar{H}&=&\omega_0 a^\dag a -J(\sigma_x \cosh ({2 \alpha 
    (a^\dag-a)})+\nonumber\\
    &&+\rmi \sigma_y \sinh ({2 \alpha (a^\dag-a)}) )+E_p/2.
\end{eqnarray}

Depending on the choice of the parameters, the problem will be better described
by a electron or  polaron excitation picture. In particular, in the weak
coupling limit, both the small  polaron and the electron are good quasiparticle
while, in the intermediate and strong coupling regimes, the  polaron behaviour
prevails \cite{robin}.

The different regimes was widely studied, in literature, both for the two site
problem and the extended case. The anti-adiabatic case was first studied in
small $J$ perturbation regime \cite{holstein,appel} and in the Holstein-Lang-Firsov
approximation \cite{holstein,lang} (HLFA), were an effective Hamiltonian is introduced to
eliminate the phonon states. In HLFA $J$ is substituted by an
effective hopping integral obtained by averaging the displacement 
$\exp{[2\alpha (a^\dag-a)]}$ on the thermal distribution of phonons. 
The resulting effective hopping integral is 
\begin{equation}\label{eqn:HLFAapproxT}
J^*=J \exp(-4\alpha^2(n_B(T)+1/2)), 
\end{equation}
where $n_B$ is the Bose occupation number. At zero temperature, the well known
exponential reduction of the bandwidth is obtained 
\begin{equation}\label{eqn:HLFAapprox}
J^*=J \exp(-2\alpha^2)
\end{equation}
and, as the temperature is increased ($T/J>>\gamma$), 
the bandwidth decreases rapidly.
As we shown in (\cite{paganelciuk}) and in the present paper, this is a good approximation at
zero temperature but it becomes inadequate at finite $T$ where incoherent
processes turns out to be important.

\section*{References}
\bibliographystyle{iopart-num}
 \bibliography{/home/simone/universita/bibliografia/bibliografia,/home/simone/universita/bibliografia/bibl-2site-coer,/home/simone/universita/bibliografia/bibl-electron-phon}

\end{document}